\documentclass[11pt]{article}
\usepackage{graphicx} 
\usepackage{feynmf}
\usepackage{amssymb}
\usepackage{amsmath}
\usepackage{xcolor}
\usepackage{cite}
\newcommand{\calc}{\text{${\cal C}$}}

\newcommand{\calf}{\text{${\cal F}$}}

\newcommand{\calt}{\text{${\cal T}$}}

\newcommand{\hr}{\hat{r}}

\newsavebox{\Gammap}
\newsavebox{\Gammam}

\begin{document}

\setlength{\unitlength}{1mm}
 \begin{fmffile}{samplepics}

\begin{titlepage}

\vspace{1.cm}

\long\def\symbolfootnote[#1]#2{\begingroup%
\def\thefootnote{\fnsymbol{footnote}}\footnote[#1]{#2}\endgroup} 

\begin{center}

%{\large \bf Sketch of a novel mixed analytical/numerical approach \\
%for the computation of two-loop $N$-point Feynman diagrams}\\[2cm]
  {\large \bf Framework for a novel mixed analytical/numerical approach \\
for the computation of two-loop $N$-point Feynman diagrams}\\[2cm]

{\large  J.~Ph.~Guillet$^{a}$, E.~Pilon$^{a}$, 
Y.~Shimizu$^{b}$ and M. S. Zidi$^{c}$ } \\[.5cm]

\normalsize
{$^{a}$ Univ. Grenoble Alpes, Univ. Savoie Mont Blanc, CNRS, LAPTH, F-74000 Annecy, France}\\
{$^{b}$ KEK, Oho 1-1, Tsukuba, Ibaraki 305-0801, Japan\symbolfootnote[2]{Y. Shimizu passed away during the completion of this series of articles.}}\\
{$^{c}$ LPTh, Universit\'e de Jijel, B.P. 98 Ouled-Aissa, 18000 Jijel, Alg\'erie}\\
      
%\vspace{1.0cm}

%\today
\end{center}

\vspace{2cm}

\begin{abstract} 
\noindent
A framework to represent and compute two-loop $N$-point 
Feynman diagrams as double-integrals is discussed. The integrands are ``generalised one-loop type'' 
multi-point functions multiplied by simple weighting factors. 
The final integrations over these two variables are to be 
performed numerically, whereas the ingredients involved in the integrands, in
particular the ``generalised one-loop type'' functions, are
computed analytically. The idea is illustrated on a few examples of scalar 
three- and four-point functions.   
\end{abstract}

\vspace{1cm}

\begin{flushright}
arXiv: 1905.08115\\
LAPTH-029/19\\
\end{flushright}

\end{titlepage}

\section{Introduction}\label{s1}

A key ingredient in an automated evaluation of two-loop multileg processes is a
fast and numerically stable evaluation of scalar Feynman integrals. The 
derivation of a fully analytic result remains beyond reach so far in the general 
mass case. On the opposite side, in particular for the calculation of two-loop 
three- and four-point functions in the general complex mass case relying on 
multidimensional numerical integration by means of sector decomposition 
\cite{Borowka:2015mxa,Borowka:2012yc,Soper:1999xk,Bogner:2007cr,Smirnov:2008py}
a reliable result has a high computing cost. Approaches based on
Mellin-Barnes techniques \cite{Czakon:2005rk,Gluza:2007rt,Smirnov:2009up,Freitas:2010nx,Gluza:2016fwh} allow to perform part of the integrals 
analytically, yet, as far as we know, the number of integrals left over 
for numerical quadratures depends on the topologies considered and can remain 
rather costly. It would therefore be useful to perform part of the Feynman
parameter integrations analytically in a systematic way to reduce the
number of numerical quadratures. 
 
\vspace{0.3cm}

\noindent
This article aims at initiating such a working program, 
advocating the implementation of two-loop
% scalar 
$N$-point 
%functions in four dimensions $^{(2)}I_{N}^{4}$ as (weighted sums of) 
functions in $n$ dimensions $^{(2)}I_{N}^{n}$ as (weighted sums of) 
double integrals in the form:
%\begin{align}
%^{(2)}I_{N}^{4}
%& \sim 
%\sum \int_{0}^{1} d \rho \int_{0}^{1} d \xi \, W(\rho,\xi) \;
%^{(1)}\widetilde{I}_{N^{\prime}}^{4}(\rho,\xi)
%\label{e-zero}
%\end{align}
\begin{align}
^{(2)}I_{N}^{n}
& \sim 
\sum \int_{0}^{1} d \rho \int_{0}^{1} d \xi \, W(\rho,\xi) \;
^{(1)}\widetilde{I}_{N^{\prime}}^{n^{\prime}}(\rho,\xi)
\label{e-zero}
\end{align}
where $W(\rho,\xi)$ are weighting functions given analytically. 
%The factors $^{(1)}\widetilde{I}_{N^{\prime}}^{4}(\rho,\xi)$  
The factors $^{(1)}\widetilde{I}_{N^{\prime}}^{n^{\prime}}(\rho,\xi)$  
are $N^{\prime}$-point functions
of some ``generalised one-loop type" in a sense explained below. 
Once the 
% the $\caln^{\prime}$-point functions 
%$^{(1)}\widetilde{I}_{N^{\prime}}^{4}(\rho,\xi)$ computed analytically, 
$^{(1)}\widetilde{I}_{N^{\prime}}^{n^{\prime}}(\rho,\xi)$ computed analytically, 
the 
%two-loop $\caln$-point functions 
%$^{(2)}I_{N}^{4}$ are obtained by numerical quadrature 
$^{(2)}I_{N}^{n}$ are obtained by numerical quadrature 
over the sole two remaining variables $\rho$ and $\xi$, which represents a 
substantial gain w.r.t.\ a fully numerical integration over the many Feynman 
parameter of the primary two-loop integral.

\vspace{0.3cm}

\noindent
In this article, we first provide a general argument in sec.~\ref{argument-general}. 
In sec. 3, we show that the polytopes spanned by the Feynman parameters related to the ``generalised one-loop functions'' can always 
be partitioned into simplices. We then illustrate the general argument given in sec. 2 considering an example of three-point scalar 
diagram $^{(2)}I_{3}^{4}$ with a non planar topology in sec.~\ref{sec3pt}, and an example of four-point scalar 
diagram $^{(2)}I_{4}^{4}$ with a non planar topology in sec.~\ref{sec4pt}
\footnote{Since the two examples studied do not have UV divergences nor IR/collinear ones, the space-time dimension $n$ is taken to $4$ for these two cases.}. 
Sec.~\ref{outlook} concludes this article with an overview of extensions of the 
present program to be presented in subsequent publications.
%Finally, Appendix~\ref{proof} provides a proof that, for any 
%general $N$-point two-loop diagram with a non planar topology, a shrewd choice 
%of parametrisation can always be found which leads to simplified building 
%blocks in the ``generalised one-loop" amplitude.
Finally, Appendix~\ref{proof} provides a proof that, for any 
general $N$-point two-loop diagram in a scalar theory with three-leg vertices with different topologies, a shrewd choice 
of parametrisation can always be found which leads to simplified building 
blocks in the ``generalised one-loop" amplitude.

\section{General argument}\label{argument-general}

Let us consider an arbitrary two-loop Feynman diagram with topology $\cal T$ 
involving $N$ external legs with external momenta $\{p_{i}, i=1, \cdots, N\}$ 
and $I$ internal lines with internal masses $\{m_{k}^2, k=1, \cdots, I\}$.
To simplify we stick here to a scalar function i.e. we ignore
sophistications that may arise from spin-carrying internal lines and/or 
derivative couplings. We need not specify the type of scalar vertices 
considered either. The integral representation of the diagram is given by: 
\begin{align}
^{(2)}I_{N}^{n} \left( \{p_{i}\};\{m_{k}^2\}, \calt \right)
\quad  
&= 
\int \left[ \prod_{l=1}^{2} \frac{d^n k_l}{(2 \pi)^n} \right]
\prod_{k=1}^{I} \frac{1}{q_k^2 - m_k^2 + i \lambda}
\label{eqG00} 
\end{align}
where the internal momenta $\{q_k\}$ are graded sums of the two loop momenta 
$k_l,\, l=1,2 $ and the external momenta $\{p_{i}\}$. In order to introduce our notations we recast eq. (\ref{eqG00}) into the 
following mixed parametric representation of this diagram:
\begin{align}
^{(2)}I_{N}^{n} \left( \{p_{i}\};\{m_{k}^2\}, \calt \right)
\quad  
&= (-i)^I \,
\int_{(I\!\!R^{+})^{I}} \left[ \prod_{k=1}^{I} d \tau_{k} \right] \, 
\delta \left( 1 - \sum_{l=1}^{I} \tau_{ \, l} \right)
\int_{0}^{+ \infty} d \alpha \, \alpha^{I-1} \,
\notag\\
& 
\quad{} \quad{}   \quad{} \quad{} 
\times
%\int_{0}^{+ \infty} d \alpha \, \alpha^{I-1} \,
\int \left[ \prod_{l=1}^{2} \frac{d^n k_l}{(2 \pi)^n} \right]
%e^{
%\left\{
\exp
\left\{ 
 i \, \alpha \, 
 \left[ 
  \sum_{k=1}^{I} \tau_{k} \left( q_k^2 - m_k^2 + i \lambda \right)
 \right]
\right\}
%}
\label{eqG0} 
\end{align}
The integration over the two loop momenta $k_i$ is made easier rewriting the 
denominator in the integrand as follows:
\begin{align}
\sum_{k=1}^{I} \tau_{k} \left( q_k^2 - m_k^2 \right)
& = 
\begin{array}{c}
 [k_1  \quad k_2] 
%\\
%\mbox{}
\end{array}
\!\! \cdot A \cdot 
 \left[
  \begin{array}{c}
    k_1 \\
    k_2
  \end{array}
 \right]
+ 2 \, 
\begin{array}{c}
 [r_1  \quad  r_2] 
%\\
% \mbox{}
\end{array}
\!\! \cdot 
 \left[
  \begin{array}{c}
   k_1 \\
   k_2
  \end{array}
 \right] 
+ \calc
  \label{eqG3}
\end{align}
In eq. (\ref{eqG3}) the elements of the $2 \times 2$ symmetric matrix $A$ are 
sums of Feynman parameters $\tau_k$'s, whereas the $n$-vectors $r_l$ are linear
combinations of external momenta $\{p_i\}$ weighted by Feynman parameters 
$\{\tau_k\}$. The compact notations mean:
\begin{align}
\begin{array}{c}
 [k_1 \quad  k_2] 
%\\
% \mbox{}
\end{array}
\!\! \cdot A \cdot 
 \left[
  \begin{array}{c}
    k_1 \\
    k_2
  \end{array}
 \right]
& = 
\sum_{j,l=1}^{2}
A_{jl} \, (k_j \cdot k_l)
\notag\\
   \begin{array}{c}
 [r_1   \quad  r_2] 
%\\
% \mbox{}
\end{array}
\!\! \cdot 
 \left[
  \begin{array}{c}
   k_1 \\
   k_2
  \end{array}
 \right] 
& = 
\sum_{l=1}^{2} (r_l \cdot k_l)
\notag
\end{align}
The term $\calc$ is of the form
\begin{align}
\calc & = 
\sum_{i,j} {\cal Q}_{ij}( p_i \cdot p_j) - \sum_{j=1}^{I} \tau_j \, m_j^2 
\notag
\end{align}
where the matrix ${\cal Q}$ is linear in the Feynman parameters $\{\tau_k\}$.
The integrations over the two loop momenta $k_l$, then over the parameter 
$\alpha$ yield\footnote{up to a constant factor 
$(-1)^{I+1} \, (4 \pi)^{-n} \, \Gamma(I-n)$  
%e^{i \frac{\pi}{2} (4-n)} \,
irrelevant here, which will be dropped in the following.}:
\begin{align}
^{(2)}I_{N}^{n} \left( \{p_{j}\}; \calt \right)
\quad  
&= 
\int_{(I\!\!R^{+})^{I}} \left[ \prod_{k=1}^{I} d \tau_{k} \right] \, 
\delta \left( 1 - \sum_{l=1}^{I} \tau_{ \, l} \right) \, 
%\notag\\
%&
%\quad \quad \quad \quad \quad
%\times 
\left[ \det (A) \right]^{- \frac{n}{2}} 
\left[ 
 {\cal D}(\{\tau_{k}\}) - i \, \lambda 
\right]^{n-I}
\label{eqG1} 
\end{align}
where the term ${\cal D}$ is given by:
\begin{align}
{\cal D}(\{\tau_{k}\})
&=  
\left\{ \sum_{i,j=1}^{2}[A^{-1}]_{ij} \, (r_i \cdot r_j) \right\}- {\calc}
\label{eqG2}
\end{align}
The determinant $\det(A)$ is real and non negative as will be seen below, 
which allows to rewrite for later convenience:
\begin{align}
\left[ \det (A) \right]^{- \frac{n}{2}} 
\left[ 
 {\cal D}(\{\tau_{k}\}) - i \, \lambda 
\right]^{n-I}
& =
\left[ \det (A) \right]^{I - \frac{3}{2} \, n} 
\left[ 
 {\cal F}(\{\tau_{k}\}) - i \, \lambda 
\right]^{n-I}
\label{eqG1bis}
\end{align}
with 
\begin{align}
{\cal F}(\{\tau_{k}\}) 
& = 
\left\{ \sum_{i,j=1}^{2} \mbox{Cof}[A]_{ij} \, (r_i \cdot r_j) \right\} 
- \det (A) \, {\calc}
\label{eqG2bis}
\end{align}
where $\mbox{Cof}[A]$ is the matrix of cofactors of $A$.
The matrix $A$, the momenta $r_i$ and the 
scalar function $\calc$ depend linearly on the $\tau_k$'s, thus $\calf$ is
homogeneous of degree 3 in the $\tau_k$'s. Besides its dependence on the 
$\{\tau_{k}\}$, the factor $\calf$ also depends
on the external momenta $\{p_{j}\}$, the internal masses $\{m_{k}^2\}$ and the
topology $\cal T$ of the diagram; these extra dependences will not be made
explicit in what follows to lighten the notations. 
The parametric representation for $^{(2)}I_{N}^{n}$ which we thereby obtained
identifies with the one introduced e.g. in \cite{eden2002analytic}. It is synthesised in:
\begin{align}
^{(2)}I_{N}^{n} \left( \{p_{j}\}; \calt \right)
\quad  
&= 
\int_{(I\!\!R^{+})^{I}} \left[ \prod_{k=1}^{I} d \tau_{k} \right] \, 
\delta \left( 1 - \sum_{l=1}^{I} \tau_{ \, l} \right) \, 
%\notag\\
%&
%\quad \quad \quad \quad \quad
%\times 
\left[ 
  \det (A)
\right] ^{I - \, \frac{3}{2} \, n} \, 
\left[ 
 {\cal F}(\{\tau_{k}\}) - i \, \lambda 
\right]^{n-I}
\label{eqG1final} 
\end{align}
The parametric representation (\ref{eqG1final}) is the actual starting point of 
this article. 

\vspace{0.3cm}

\noindent
At this stage, we may note that spin-carrying internal lines and/or derivative 
couplings would amount to some Feynman parameter-dependent numerator together
with modifications of the powers which the factors $\det(A)$ and $\calf$ in eq.
(\ref{eqG1}) are raised to.  
Yet these sophistications, together with the combinatoric relations relating 
the number of external and internal lines and of vertices of the various kinds, 
which would come from the specification of the types of particles and vertices
involved, are beside the point which we wish to make here.

\vspace{0.3cm}

\noindent
Let us partition the set 
of Feynman parameter labels $\{1, \cdots,I\}$ into three subsets $S_j$ and 
define three auxiliary parameters $\rho_j$, $j=1,2,3$ accordingly as follows:
i) $S_1$ contains the labels of the internal lines involving only $k_1$ not 
$k_2$, to $S_1$ is associated $\rho_1 \equiv \sum_{i \in S_1} \tau_i$;
ii) $S_2$ contains the labels of internal lines involving only $k_2$ not $k_1$, 
to $S_2$ is associated $\rho_2 \equiv \sum_{i \in S_2} \tau_i$;
iii) $S_3$ contains the labels of internal lines common to the two overlapping 
loops. {\em Each} of these lines involves 
the {\em same} combination\footnote{It could alternatively involve $k_1-k_2$ 
{\em in every internal line common to the two overlapping 
loops}, depending on the convention adopted for the orientations of the loop 
momenta.} $k_1+k_2$, so that the matrix element $A_{12}$ weighting the scalar 
product $(k_1 \cdot k_2)$ in the first term of eq. (\ref{eqG3}) is equal to 
the combination $\rho_3 \equiv \sum_{i \in S_3} \tau_i$. 
The $\rho_j$'s thus fulfil the constraint 
\begin{equation}
\rho_1+\rho_2+\rho_3= \sum_{j=1}^{I} \tau_j = 1
\label{eqG5}
\end{equation}
The elements of the matrix $A$ read:
\begin{equation}
A_{12} =\rho_3, 
\quad 
A_{11} = \rho_1 + \rho_3,
\quad \text{and} \quad 
A_{22} = \rho_2 + \rho_3
\label{eqG4}
\end{equation}
Hence: 
\begin{align}
\det(A) = \rho_1 \, \rho_2 + \rho_2 \, \rho_3 + \rho_3 \, \rho_1
\label{eqG5bis}
\end{align}
The determinant $\det(A)$ is clearly non negative.
Let $|S_j|$ be the number of elements of $S_j$, with $|S_1|+|S_2|+|S_3| = I$. 
Let us introduce $|S_j|$ parameters $u_{k_j}$ with 
$k_j \in S_j$ so as to reparametrise the 
$\tau_{k_j}$ summing up into $\rho_j$ as follows:
\begin{align}
\tau_{k_j} 
&= \rho_j \, u_{k_j}
\quad \text{with the constraint} \quad
\sum_{k_j \in S_j} u_{k_j} = 1
\label{eqG7}
\end{align}
Accordingly the reparametrised integration measure takes the 
following factorised form:
\begin{align}
&\left[ \prod_{j=1}^{I} d \tau_{j} \right] \, 
\delta \left( 1 - \sum_{i=1}^{I} \tau_i \right) \notag\\ 
&\quad \quad \quad 
= 
\prod_{k=1}^{3} 
\left\{ 
 d \rho_k \rho_k^{|S_k|-1}  \, 
 \prod_{j_k \in S_k} 
 \left[ 
  du_{j_k} \, \delta \left( 1 - \sum_{l \in S_k} u_l \right) 
 \right]
\right\} \, 
\delta \left( 1 - \sum_{i=1}^{3} \rho_i \right) \,
\label{jacob3}
\end{align}
With this reparametrisation, the elements of the $A$ matrix depend
only on the parameters $\rho_j$ and on none of the $u_i$'s, so do
$\text{Cof}[A]$ and $\det(A)$.  In $\calf$, the dependence in the $u_i$'s enters
through the factors  $(r_i \cdot r_j)$, quadratically, and through the term
${\cal C}$, linearly.  The term $\calf$  may thus be seen as a polynomial of
second degree in the $u_i$'s and can thus be  interpreted as building up the
integrand of a ``generalised" one-loop function  represented as a Feynman
integral over the $u_i$'s. The integral representation of the two-loop diagram 
$^{(2)}I_{N}^{n} \left( \{p_{j}\}; \calt \right)$ can thus be recast in
the following form:
\begin{align}
^{(2)}I_{N}^{n} \left( \{p_{j}\}; \calt \right)
&= 
\int_{(I\!\!R^{+})^{3}} 
\left[ \prod_{k=1}^{3} d \rho_{k} \, \rho_k^{|S_k|-1} \right] 
\delta \left( 1 - \sum_{l=1}^{3} \rho_{ \, l} \right) \, 
\left[ 
  \rho_1 \, \rho_2 + \rho_2 \, \rho_3 + \rho_3 \, \rho_1
\right] ^{I - \, \frac{3 \, n}{2}} \;
^{(1)}\widetilde{I}_{N^{\prime}}^{n^{\prime}}
\label{eqG1a}
\end{align}
%with the $N^\prime$-point function of ``generalised one-loop type" given by:
where we have introduced 
\begin{align}
^{(1)}\widetilde{I}_{N^{\prime}}^{n^{\prime}}
& =
%\int_{(I\!\!R^{+})^{I-3}} 
\int_{(I\!\!R^{+})^{I}} 
\prod_{k=1}^{3} 
\prod_{j \in S_k} d u_{j} \, \delta \left( 1 - \sum_{l \in S_k} u_{j} \right)
\left[ \,
 \overline{\cal F}(\{u_{k}\},\{\rho_l\}) - i \, \lambda 
\right]^{n-I}
\label{eqG1b}
\end{align}
with
$\overline{\cal F}(\{u_{k}\},\{\rho_l\})
= 
{\cal F}(\{\tau_{i}(\{u_{k}\},\{\rho_l\})\})
$ and we have set $N^{\prime} = I - 2$ and $ n^{\prime} = 2 \, (n - 2)$.
%so that, in eq. (\ref{eqG1b}), $n-I = - N^{\prime} + n^{\prime}/2$
%and $^{(1)}\widetilde{I}_{N^{\prime}}^{n^{\prime}}$ can thus be seen as a 
%$N^\prime$-point function of ``generalised one-loop type" in $n^{\prime}$
%dimensions.
%%Futhermore, $N^{\prime} = I - 2$ and $ n^{\prime} = 2 \, n - 4$. 
The reparametrisation of 
$^{(2)}I_{N}^{n} \left( \{p_{j}\}; \calt \right)$ according to eqs.
(\ref{eqG1a}), (\ref{eqG1b}) has already been used in the literature
\cite{Fujimoto:1995ev,Kurihara:2005ja,Yuasa:2011ff,deDoncker:2012fro} in order
to perform the integration over all Feynman parameters fully numerically.
We alternatively wish to advocate here the separate identification of 
$^{(1)}\widetilde{I}_{N^{\prime}}^{n^{\prime}}$ in eq. 
(\ref{eqG1b}) with $n-I = - N^{\prime} + n^{\prime}/2$ as a 
$N^\prime$-point function of ``generalised one-loop type" in $n^{\prime}$
dimensions, and the possibility to compute 
$^{(1)}\widetilde{I}_{N^{\prime}}^{n^{\prime}}${\it analytically}.

\vspace{0.3cm}

\noindent
The above qualificative ``generalised one-loop type" refers to two kinds of 
generalisations. \\ 
{\bf 1)} After integrating over three of the $u_i$'s in order to eliminate 
the $\delta(1 - \sum_{l \in S_k} u_{j})$-constraints, the effective kinematics 
of the ``generalised" one-loop $N^\prime$-point function in $n^\prime$
dimensions is encoded in a $(I-3) \times (I-3)$ matrix 
$G = G(\{p_j\},\{\rho_l\})$, 
a column $(I-3)$-vector $V = V(\{p_j\},\{\rho_l\})$ and a scalar function 
$C = (\{p_j\},\{\rho_l\})$, all of which functions of the external momenta 
$\{p_j\}$ and of the integration variables $\{\rho_k\}$ seen as external 
parameters. The matrix $G$ somehow plays the role of an ``effective 
Gram matrix", with which it shares a few features, namely it is real and 
symmetric and it does not depend on the (possibly complex) internal masses. 
Although not made explicit, $V$ and $C$ depend on the internal masses. 
Let us note that this effective 
kinematics of the ``generalised" one-loop function depends 
on the $\rho_j$ seen as ``external" parameters beside the external momenta 
$p_k$'s, and that it may span a larger parameter space than the one involved 
in standard one-loop $N^\prime$-point functions involved in collider 
processes at one loop. \\
{\bf 2)} Unlike for the standard one-loop function, the integration domain of 
the parameters $u_k$'s is not the usual $(I-3)$-simplex defined by 
$\Sigma_{(I-3)} = \{ u_k \geq 0, k=1, \cdots, I-3 | \sum_{k=1}^{I-3} u_k =1\}$ 
but instead the polysimplicial
%As the partitionning of the set of internal lines indices, t
set\footnote{The polysimplicial set depends on the topology 
$\cal T$ of the two-loop diagram considered. It is understood that, 
in case some of the $|S_j|$ equals 1, 
the corresponding trivial set factor $\Sigma_{(|S_j|-1)}$ shall be omitted.} 
$\Sigma_{(|S_1|-1)} \times \Sigma_{(|S_2|-1)} \times \Sigma_{(|S_3|-1)}$;
The quantity $\overline{\calf}$ formally reads:
\begin{align}
\overline{\calf} 
&= 
U^T \cdot G \cdot U - 2 \; V^T \cdot U - C
\label{eqG8c} 
\end{align}
where $U$ is the column $(I-3)$-vector gathering the yet unintegrated $(I-3)$ 
variables $u_k$ parametrising the polysimplicial integration domain 
$\Sigma_{(|S_1|-1)} \times \Sigma_{(|S_2|-1)} \times \Sigma_{(|S_3|-1)}$. 
The two-dimensional integral representation and the corresponding weighting
function $W$ advocated in eq. (\ref{e-zero}) are readily obtained from eq. 
(\ref{eqG1a}) using a reparametrisation of the form 
$\rho_1 = \rho \,\xi$, $\rho_2 = \rho \, (1-\xi)$, $\rho_3= (1-\rho)$ with 
$0 \leq \rho, \xi \leq 1$.
%A few illustrative examples are provided in the next two sections. 
A few illustrative examples are provided in the sections \ref{sec3pt} and \ref{sec4pt}. 
More details will be provided in subsequent papers.

\section{Decomposition of polytopes into simplices}\label{dec_simp}

Since the integration domain spanned by the $u_i$'s is not the 
simplex anymore, one might fear that the above functions of ``generalised 
one-loop type'' be not computable analytically because the standard one-loop 
methods could not be used. This fear is groundless, as general theorems 
on triangulation \cite{Goodman:1997} ensure that any $(I-3)$-polytope can be 
partitioned into $(I-3)$-simplices; furthermore 
the polytope ${\cal V} =\Sigma_{|S_{1}|-1} \times 
\Sigma_{|S_{2}|-1} \times \Sigma_{|S_{3}|-1}$ spanned by the $u_i$'s being
convex, the imaginary part ${\cal I}m(\overline{\cal F} - i \lambda)$ keeps a
constant (minus) sign over ${\cal V}$ even with internal general complex 
masses, hence it does also over each of the simplices involved in such a 
partition.

\subsection{Two-loop three-point functions with three-leg vertices}

\begin{figure}[h]
\begin{center}
  \includegraphics[scale=0.5]{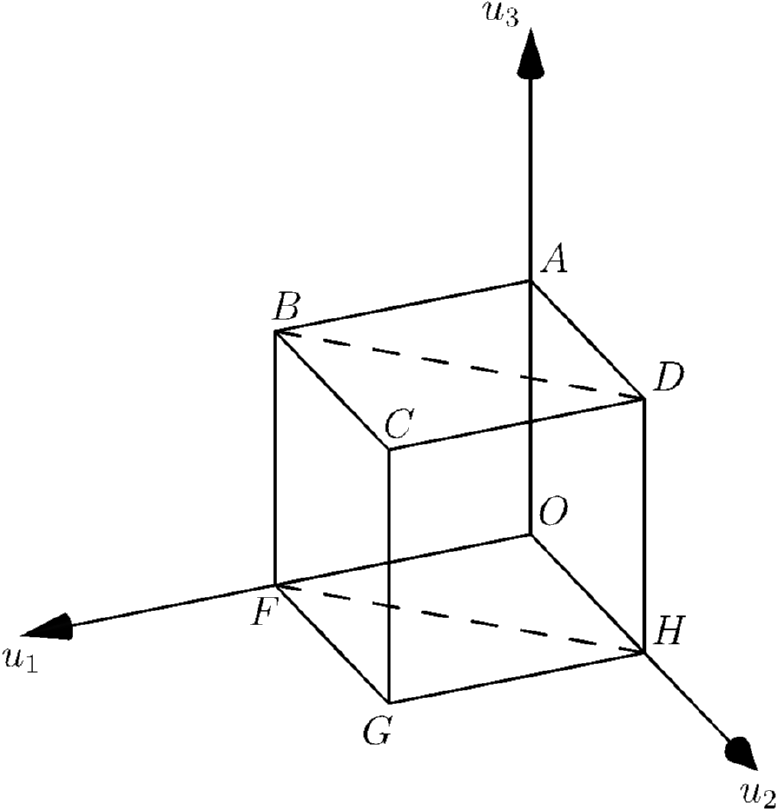}
\end{center}
\caption{\footnotesize Partitioning of the cube $K_{(3)}$ into the two wedges
$\Pi_{(3)}^{(0)}$ and $\Pi_{(3)}^{(1)}$ by the plane passing through
$B,D,H,F$.}\label{cube}
\end{figure}

In this section, we explicitly build such partitions in the cases of
two-loop three-point functions involving three-leg vertices. 
In these cases, ${\cal V}$ being three-dimensional, such a decomposition can 
thus be readily performed and visualised. Anticipating on the next 
section, the domains spanned by the $u_i$'s for, respectively, the planar 
and non planar two-loop three-point
functions are the wedge 
$\Pi_{(3)}^{(0)} = \{0 \leq u_1,u_2,u_3 \leq 1; u_1 + u_2 \leq 1\}$ 
and the cube $K_{(3)} = \{ 0 \leq u_1,u_2,u_3 \leq 1\}$. The cube
$K_{(3)}$, depicted in fig. \ref{cube}, can be partitioned into the two 
wedges: $\Pi_{(3)}^{(1)} = \{0 \leq u_1,u_2,u_3 \leq 1; u_1 + u_2 \geq 1\}$ 
and  $\Pi_{(3)}^{(0)}$ by the plane $\{u_1+u_2=1\}$. 
The mapping $(u_1,u_2,u_3) \rightarrow (1-u_2,1-u_1,u_3)$ exchanges 
$\Pi_{(3)}^{(1)}$ and $\Pi_{(3)}^{(0)}$.

\vspace{0.3cm}

\noindent
Let us focus on $\Pi_{(3)}^{(0)}$ and partition it into the 
simplex $\Sigma_{(3)} = \{0 \leq u_1,u_2,u_3 \leq 1; u_1 + u_2 + u_3 \leq 1\}$ 
and the leftover volume denoted by $\Xi_{(3)}$. 
The latter can be partitioned into the following two tetrahedra:
$\Theta_{D}^{ABH} = \{ 0 \leq u_1,u_2,u_3 \leq 1; u_1+u_2 \leq 1; u_2+u_3 \geq
1 \}$ and $\Theta_{F}^{ABH} = \{0 \leq u_1,u_2,u_3 \leq 1; u_1+u_2 \leq 1;
u_2+u_3 \leq 1; u_1+u_2+u_3 \geq 1 \}$, by the plane $\{u_2+u_3 = 1\}$. These
two tetrahedra are not isometric to $\Sigma_{(3)}$ because, for example:
\begin{equation}
  \left.
  \begin{array}{ccccccc}
    l[DA] & = & l[DH] & = & l[AB] & = & 1 \\
    l[AH] & = & l[DB] & & & = & \sqrt{2} \\
    l[BH] & & & & &= & \sqrt{3}
  \end{array}
\right\} \quad \Theta_{D}^{ABH}
  \label{eqdisttetra}
\end{equation}
where $l[AB]$ is the distance between the points $A$ and $B$. Nevertheless
these two tetrahedra can be mapped onto $\Sigma_{(3)}$ by affine 
transformations.
Indeed, for any point $M$ in the tetrahedron $\Theta_{D}^{ABH}$, the vector
$\overrightarrow{OM}$ can be written as:
\begin{equation}
  \overrightarrow{OM} = \overrightarrow{OD} + \overrightarrow{DM}
  \label{eqsplitvec}
\end{equation}
Let us define the following basis of unit vectors : 
$\vec{e}_1 =\overrightarrow{OF}$, 
$\vec{e}_2 = \overrightarrow{OH}$ and 
$\vec{e}_3 = \overrightarrow{OA}$. 
The vectors $\overrightarrow{OM}$ and $\overrightarrow{OD}$ read 
$\overrightarrow{OM} = \sum_{i=1}^{3} u_i \, \vec{e}_i$ and 
$\overrightarrow{OD} = \vec{e}_2 + \vec{e}_3 \equiv u_{Di} \, \vec{e}_i$.
The vector $\overrightarrow{DM}$ can be written $\overrightarrow{DM} =
\sum_{i=1}^{3} y_i \, \vec{a}_i$ with 
$\vec{a}_1 \equiv \overrightarrow{DA} = -\vec{e}_2$, 
$\vec{a}_2 \equiv \overrightarrow{DB} = \vec{e}_1 - \vec{e}_2$ and
$\vec{a}_3 \equiv \overrightarrow{DH} = - \vec{e}_3$. 
Let us define the $3 \times 3$ matrix $N$ such that 
$\vec{a}_i = \sum_{j=1}^{3} N_{ij} \, \vec{e}_j$. In terms of the $u_i$'s 
eq. (\ref{eqsplitvec}) translates into:
\begin{align}
  U &= U_D + N^T \cdot Y
  \label{eqlienXY}
\end{align}
with:
\begin{equation}
  U = \left(
  \begin{array}{c}
    u_1 \\
    u_2 \\
    u_3
  \end{array}
\right) 
\quad , \quad
  U_D = \left(
  \begin{array}{c}
    0 \\
    1 \\
    1
  \end{array}
\right) 
\quad , \quad
  Y = \left(
  \begin{array}{c}
    y_1 \\
    y_2 \\
    y_3
  \end{array}
\right) 
\quad , \quad 
  N = \left(
  \begin{array}{ccc}
    0 & -1 & 0 \\
    1 & -1 & 0 \\
    0 & 0 & -1
  \end{array}
\right)
  \label{eqmatN}
\end{equation}
Using eq. (\ref{eqlienXY}), the constraints on the $u_i$'s of a point
spanning the tetrahedron $\Theta_{D}^{ABH}$, which read
$\{ 0 \leq u_1,u_2,u_3
\leq 1; u_1+u_2 \leq 1; u_2+u_3 \geq 1 \}$, 
translate into 
$\{0 \leq y_1,y_2,y_3 \leq 1; 0 \leq y_1 + y_2 + y_3 \leq 1 \}$ in terms of 
the $y_i$'s i.e. $Y$ spans $\Sigma_{(3)}$. 
Likewise, for the other tetrahedron $\Theta_{F}^{ABH}$
the constraints read 
$\{0 \leq u_1,u_2,u_3 \leq 1; u_1+u_2 \leq 1; u_2+u_3 \leq 1;
u_1+u_2+u_3 \geq 1 \}$. Performing the reparametrisation
\begin{equation}
  \left(
  \begin{array}{c}
    u_1 \\
    u_2 \\
    u_3
  \end{array}
\right) = 
\left(
\begin{array}{c}
  1 - y^{\prime}_1 - y^{\prime}_2 \\
  y^{\prime}_2 \\
  1 - y^{\prime}_2 - y^{\prime}_3
\end{array}
\right)
  \label{eqtransfthetaf}
\end{equation}
the column vector $Y^{\prime}$ is readily seen to span $\Sigma_{(3)}$ too. 

\begin{figure}[h]
\begin{center}
  \includegraphics[scale=0.5]{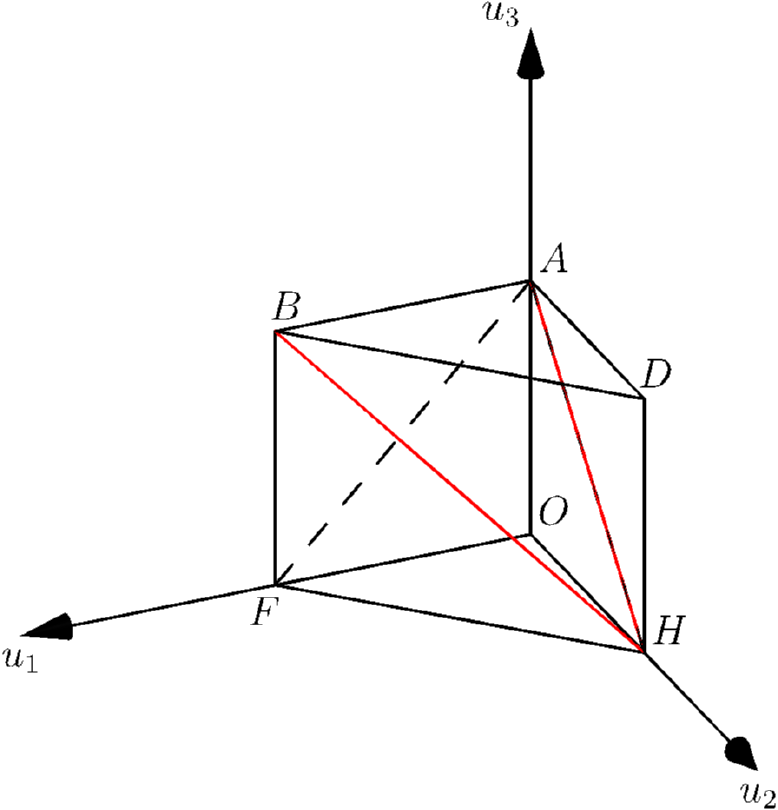}
\end{center}
\caption{\footnotesize Decomposition of the wedge $\Pi_{(3)}^{(0)}$ into the
simplex $\Sigma_{(3)}$ and two tetrahedra $\Theta_{D}^{ABH}$ and 
$\Theta_{F}^{ABH}$.}\label{wedge}
\end{figure}

\noindent
Under an affine transformation of the type (\ref{eqlienXY}): 
$U = U_0 + N^T \cdot Y$, the second order polynomial $\overline{\calf}(U)$ can
be rewritten as:
\begin{align}
  \overline{\calf}(U) &= U^T \cdot G \cdot U + 2 \, V^T \cdot U + C \notag \\
  &= Y^T \cdot ( N \cdot G \cdot N^T ) \cdot Y + 2 \, [ (U_0^T \cdot G + V^T) \cdot N^T ] \cdot Y + U_0^T \cdot G \cdot U_0 + 2 \, V^T \cdot U_0 + C \notag \\
  &\equiv Y^T \cdot G^{\prime} \cdot Y + 2 \, V^{\prime \, T} \cdot Y + C^{\prime}
  \label{eqcalftransf}
\end{align}
The new variables span $\Sigma_{(3)}$ in all contributions of the partition, 
and the second order polynomials take similar forms yet with Gram matrices
$G^{(\prime)}$, vectors $V^{(\prime)}$ and constants $C^{(\prime)}$ 
depending on the parameters of the reparametrisations $\{U_0,N\}$ and proper 
to each contribution.

\subsection{Other types of two-loop functions}

For two-loop four point functions with three-leg vertices, and other cases for
which ${\cal V}$ is four-dimensional or more, things are harder to visualise.
Nevertheless, as already mentioned general theorems \cite{Goodman:1997} on 
triangulation ensure that  ${\cal V}$ can be partitioned into simplicies 
---~this is the higher dimensional generalisation of the property whereby any 
polygon can be partitioned into triangles~--- and each of these 
$(I-3)$-simplices can be further mapped onto the normalised $(I-3)$-simplex 
$\Sigma_{(I-3)}$ by an affine transformation of the $u_i$ coordinates. 
This proves that standard methods used to analytically compute ordinary 
one-loop $N^{\prime}$-point functions can also be used to analytically compute 
$N^{\prime}$-point functions of ``generalised one-loop type"\footnote{Strictly speaking, 
for UV or IR/collinear divergent cases, the expansion around $n^{\prime}-4$ of the standard 
method for one-loop computations has to be pushed further in order to keep all the relevant terms.}
. So long for the principles.

\vspace{0.3cm}

\noindent
From a practical point of view, several triangulation methods are available, 
and the number of simplices as well as their sizes and shapes depend on the
method used. The integration over each simplex will generate dilogarithms,
some of which will cancel against other dilogarithms coming from other 
simplicies. Accordingly a question arises and practically matters regarding 
computational efficiency: what is the method which optimises the number of 
simplices so as to minimise the number of dilogarithms? 

\vspace{0.3cm}

\noindent
We will not elaborate more on this point in this article because we 
tackled the analytic computation and the issue of taming the proliferation 
of dilogarithms in an alternative way. In ref. \cite{paper1} we have developed 
an alternative method to integrate over the 
$u_i$'s which makes the integration over each Feynman parameter $u_i$ trivial, 
by representing the integrand as a derivative with respect to this Feynman 
parameter. For this purpose we use a ``Stokes-type'' identity and an integral 
representation requiring the introduction of an extra variable ranging from 
$0$ to $+\infty$ for each $u_i$. This handling is iterated until no more 
Feynman parameter remains to be integrated over. These extra integral 
representations are then easily undone later on~--- this trick plays a role 
similar to a ``catalyst" in chemistry.\\ 

\noindent
The domain $[0,+\infty[^{I-3}$ on which the auxiliary variables are 
integrated over is independent of the initial volume spanned by the 
$u_i$'s. For example, for a four-point``generalised one-loop function'' it is 
the principal octant. Thus, for a given number of external legs, whatever the 
volume spanned by the $u_i$'s is, the
domain of the leftover integrations over the new variables are always the same
as in the case of the genuine one-loop, only the number of terms to integrate
differs, and for all these terms the integration is of the same type. In
other words, the method developed in ref. \cite{paper1} and applied to the 
usual one-loop case where the volume of integration is a simplex merely for 
sake of illustration, can be used for more general integration domains, the 
only change will be the number of terms in the integrand. This alternative
method thus provides another practical way to analytically compute the
``generalised one-loop functions''. The explicit application of this method to the
computation of the ``generalised one-loop functions''
and its optimisation to tame the number of dilogarithms generated will be
further elaborated in subsequent publications.

\section{An example of scalar two-loop three-point topology}\label{sec3pt}

\begin{figure}[h]
\centering
\parbox[c][53mm][t]{50mm}{\begin{fmfgraph*}(50,50)
  \fmfstraight
  \fmfleftn{i}{1} \fmfrightn{o}{2}
  \fmfset{arrow_len}{3mm}
  \fmf{fermion,label={\small $p_1$}}{i1,v2}
  \fmf{fermion,label={\small $p_2$}}{o1,v3}
  \fmf{fermion,label={\small $p_3$}}{o2,v4}
  \fmf{phantom,tension=0.5}{v2,v2p,v3}
  \fmf{phantom,tension=0.5}{v4,v1p,v2}
  \fmffreeze
  \fmf{fermion,label={\small $q_6$},label.side=right}{v2p,v4}
  \fmf{fermion,rubout,label={\small $q_4$},label.side=left}{v1p,v3}
  \fmf{fermion,label={\small $q_2$},label.side=right}{v2,v2p}
  \fmf{fermion,label={\small $q_3$}}{v3,v2p}
  \fmf{fermion,label={\small $q_5$},label.side=right}{v4,v1p}
  \fmf{fermion,label={\small $q_1$}}{v1p,v2}
\end{fmfgraph*}}
\caption{\small The diagram picturing the two-loop three-point function with 
a non planar topology.}
\label{fig2-1} 
\end{figure}
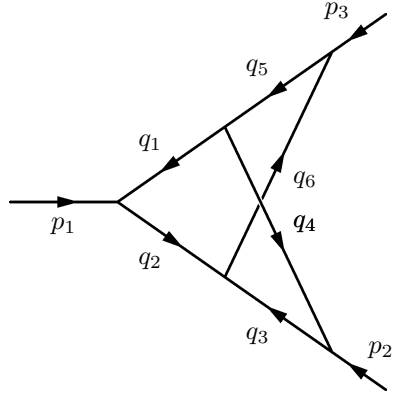

For illustrative purpose let us consider the non-planar diagram drawn on 
fig. \ref{fig2-1}. With $N=3$, $I=6$ and $n=4$ this diagram 
has the following parametric integral representation:
\begin{align}
^{(2)}I_{3}^{4} \left( \{p_{j}\}; \calt \right)
\quad  
&= 
\int_{(I\!\!R^{+})^{6}} \left[ \prod_{k=1}^{6} d \tau_{k} \right] \, 
\delta \left( 1 - \sum_{l=1}^{6} \tau_{ \, l} \right) \, 
%\notag\\
%&
%\quad \quad \quad \quad \quad
%\times 
%\left[ 
  %\det (A)
%\right] \, 
\left[ 
 {\cal F}(\{\tau_{k}\}) - i \, \lambda 
\right]^{-2}
\label{eq3pt1}
\end{align}
The factor $\calf$, whose cumbersome expression is not made explicit here,
involves the matrix\footnote{In this case --- and similarly for the planar 
topology --- $N-\frac{3}{2}n =0$ thus the factor $\det(A)$ present in eq. 
(\ref{eqG1final}) does not appear in eq. (\ref{eq3pt1}).} $A$ given by
\begin{align}
  A = \left[
  \begin{array}{cc}
     \tau_1+\tau_2+\tau_5+\tau_6 & \tau_5+\tau_6 \\
     \tau_5+\tau_6 & \tau_3+\tau_4+\tau_5+\tau_6
  \end{array}
  \right]
  \label{eq3pt2}
\end{align}
Let us reparametrise the Feynmam parameters $\tau_k$'s as follows:
\begin{align}
\rho_1 
= \tau_1+\tau_2, \quad \rho_2 = \tau_3+\tau_4 
\quad 
\text{and} 
\quad 
\rho_3 = \tau_5+\tau_6
\label{eq3pt3}
\end{align}
further with
\begin{align}
\tau_1 
&= \rho_1 \, u_1, 
& 
\tau_2 
&= \rho_1 \, (1 - u_1), 
\notag \\
\tau_3 
&= \rho_2 \, (1 - u_2), 
& 
\tau_4 
&= \rho_2 \, u_2, 
\notag \\
\tau_5 
&= \rho_3 \, u_3, 
& 
\tau_6 
&= \rho_3 \, (1 - u_3), 
\label{eq3pt4}
\end{align}

\noindent
Let us note $k_1$ and $k_2$ the four-momenta running into the loops, the internal four-momenta of fig. \ref{fig2-1} are given by:
\begin{align}
  q_1 &= k_1 - p_1, & q_2 &= k_1, & q_3 &= k_2, \notag \\
  q_4 &= k_2 - p_2, & q_5 &= k_1 + k_2 - p_1 - p_2, & q_6 &= k_1 + k_2
  \label{eq3ptmom}
\end{align}
In addition, a mass $m_i$ is associated to each internal line with four-momentum $q_i$.
Each variable $u_1$, $u_2$ and $u_3$ spans the interval $\Sigma_{(1)} =
[0,1]$.\\
The matrix $G$ reads:
\begin{align}
G_{11} 
&= \rho_1^2 \, (\rho_2+\rho_3) \, p_1^2 
& 
G_{12} 
&= \frac{1}{2} \, \rho_1 \, \rho_2 \, \rho_3 \, (p_1^2 + p_2^2 - p_3^2) 
& 
G_{13} 
&= \frac{1}{2} \, \rho_1 \, \rho_2 \, \rho_3 \, (p_3^2 - p_2^2 + p_1^2) 
\notag \\ 
G_{22} 
&= \rho_2^2 \, (\rho_1+\rho_3) \, p_2^2 
& 
G_{23} 
&= \frac{1}{2} \, \rho_1 \, \rho_2 \, \rho_3 \, (p_2^2 + p_3^2 - p_1^2) 
&
G_{33} 
&= \rho_3^2 \, (\rho_1+\rho_2) \, p_3^2
\label{eq3pt4p}
\end{align}
whereas the vector $V$ reads:
\begin{align}
V_1 
&= \frac{1}{2} \,  \rho_1 \, (\rho_1 \, \rho_2 + \rho_2 \, \rho_3 + \rho_3 \, \rho_1) \,
\left[ 
 p_1^2 + m_2^2 - m_1^2  
\right] 
\notag \\
V_2 
&=  \frac{1}{2} \, \rho_2 \, (\rho_1 \, \rho_2 + \rho_2 \, \rho_3 + \rho_3 \, \rho_1) \,
\left[ 
 p_2^2 + m_3^2 - m_4^2 
\right] 
\notag \\
V_3 
&=  \frac{1}{2} \, \rho_3 \,  (\rho_1 \, \rho_2 + \rho_2 \, \rho_3 + \rho_3 \, \rho_1) \,
\left[ 
 p_3^2 + m_6^2 - m_5^2
\right]
  \label{eq3pt5}
\end{align}
and $C$ is given by:
\begin{align}
C 
&= - (\rho_1 \, \rho_2 + \rho_2 \, \rho_3 + \rho_3 \, \rho_1) \, 
     (\rho_1 \, m_2^2 + \rho_2 \, m_3^2 + \rho_3 \, m_6^2)
\label{eq3pt5p}
\end{align}
The explicit expression for $\overline{\calf}$ then follows from eq. 
(\ref{eqG8c}). We thereby get the advocated integral representation 
\begin{align}
^{(2)}I_{3}^{4} \left( \{p_{j}\}; \calt \right)
\quad  
&= 
\int_{(I\!\!R^{+})^{3}} 
d \rho_{1} \, d \rho_{2} \, d \rho_{3} \; 
\rho_1 \, \rho_2 \, \rho_3 \; 
\delta \left( 1 - \sum_{l=1}^{3} \rho_{ \, l} \right) \; 
^{(1)}\widetilde{I}_{4}^{4}
\label{eq3pt6a}
\end{align}
involving the four-point function of ``generalised one-loop type" given by:
\begin{align}
^{(1)}\widetilde{I}_{4}^{4}
& =
\int_{0}^{1} d u_{1} \int_{0}^{1} d u_{2} \int_{0}^{1} d u_{3} 
\left[ \,
 \overline{\cal F}(\{u_{k}\},\{\rho_l\}) - i \, \lambda 
\right]^{-2}
\label{eq3pt6b}
\end{align}
The change of variables $\rho_1 = \rho \, \xi$, $\rho_2 = \rho \,(1-\xi)$,
$\rho_3=(1-\rho)$ amounts in eq. (\ref{eq3pt6a}) to the replacement
\[
\int_{(I\!\!R^{+})^{3}} 
d \rho_{1} \, d \rho_{2} \, d \rho_{3} \; 
\rho_1 \, \rho_2 \, \rho_3 \; 
\delta \left( 1 - \sum_{l=1}^{3} \rho_{ \, l} \right) \, 
 \times
\; \to \;
\int_{0}^{1} d \rho  \int_{0}^{1} d \xi \, W(\rho,\xi) \times
\]
where the weighting function $W(\rho,\xi)$ is given by:
\begin{align}
W(\rho,\xi) 
&=
\rho^3 \, (1-\rho) \, \xi \, (1-\xi) 
\label{eq3pt6c}
\end{align}
We chose to illustrate our purpose with the three-point non-planar topology 
whose corresponding integral is usually considered more touchy to compute 
than for the planar topology with the same three-leg type vertices. 
The latter can all be worked out in a very similar way, and
%. In this case in particular 
the domain of integration over the parameters $u_1,u_2,u_3$ 
%similarly introduced 
is found to be the cylinder with triangular cross section 
$\Sigma_{(2)} \times [0,1]$ where 
$\Sigma_{(2)} = \{ 0 \leq u_1,u_2,u_1+u_2 \leq 1\}$ instead of the unit cube.

\section{An example of scalar two-loop four-point topology}\label{sec4pt}

\begin{figure}[h]
\centering
\parbox[c][63mm][t]{50mm}{\begin{fmfgraph*}(60,80)
  \fmfstraight
  \fmfleftn{i}{4} \fmfrightn{o}{4}
  \fmfset{arrow_len}{3mm}
  \fmf{fermion,label={\small $p_1$}}{i3,v1}
  \fmf{fermion,label={\small $p_2$}}{i2,v2}
  \fmf{fermion,label={\small $p_3$}}{o2,v3}
  \fmf{fermion,label={\small $p_4$}}{o3,v4}
  \fmf{fermion,tension=0.2,label={\small $q_2$}}{v1,v2}
  \fmf{phantom,tension=0.5}{v2,v2p,v3}
  \fmf{phantom,tension=0.2}{v3,v4}
  \fmf{phantom,tension=0.5}{v4,v1p,v1}
  \fmffreeze
  \fmf{plain}{v2p,v6,v4}
  \fmf{fermion,label={\small $q_7$}}{v6,v4}
  \fmf{plain,rubout}{v3,v7,v1p}
  \fmf{fermion,rubout,label={\small $q_5$},label.side=left}{v7,v3}
  \fmf{fermion,label={\small $q_3$}}{v2,v2p}
  \fmf{fermion,label={\small $q_4$}}{v3,v2p}
  \fmf{fermion,label={\small $q_6$}}{v4,v1p}
  \fmf{fermion,label={\small $q_1$}}{v1p,v1}
\end{fmfgraph*}}
\caption{\small The box picturing the two-loop four-point function with 
a non planar topology.}
\label{fig3-1} 
\end{figure}
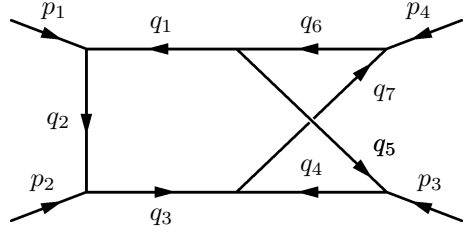

Let us now consider the non-planar diagram drawn on 
fig. \ref{fig3-1}. With $N=4$, $I=7$ and $n=4$ this diagram 
has the following parametric integral representation:
\begin{align}
^{(2)}I_{4}^{4} \left( \{p_{j}\}; \calt \right)
\quad  
&= 
\int_{(I\!\!R^{+})^{7}} \left[ \prod_{k=1}^{7} d \tau_{k} \right] \, 
\delta \left( 1 - \sum_{l=1}^{7} \tau_{ \, l} \right) \, 
%\notag\\
%&
%\quad \quad \quad \quad \quad
%\times 
\left[ 
  \det (A)
\right] \, 
\left[ 
 {\cal F}(\{\tau_{k}\}) - i \, \lambda 
\right]^{-3}
\label{eq4pt1}
\end{align}
where the matrix $A$ is given by
\begin{align}
  A = \left[
  \begin{array}{cc}
     \tau_1+\tau_2+\tau_3+\tau_6+\tau_7 & \tau_6+\tau_7 \\
     \tau_6+\tau_7 & \tau_4+\tau_5+\tau_6+\tau_7
  \end{array}
  \right]
  \label{eq4pt2}
\end{align}
The cumbersome expression of $\calf$ is not made explicit here.
Let us reparametrise the Feynmam parameters $\tau_k$'s as follows:
\begin{align}
\rho_1 
= \tau_1+\tau_2+\tau_3, \quad \rho_2 = \tau_4+\tau_5 
\quad 
\text{and} 
\quad 
\rho_3 = \tau_6+\tau_7
\label{eq4pt3}
\end{align}
further with
\begin{align}
\tau_1 
&= \rho_1 \, u_1, 
& 
\tau_2 
&= \rho_1 \, u_2,
& 
\tau_3 
&= \rho_1 \, (1 - u_1 - u_2),
\notag \\
\tau_4 
&= \rho_2 \, (1 - u_3), 
& 
\tau_5 
&= \rho_2 \, u_3, 
& 
& 
\notag \\
\tau_6 
&= \rho_3 \, u_4, 
& 
\tau_7 
&= \rho_3 \, (1-u_4)
& 
& 
\label{eq4pt4}
\end{align}
%Variables $\rho_1,\rho_2,\rho_3$ span the 2-simplex 
%$\Sigma_{(2)} = \{0 \leq \rho_1,\rho_2,\rho_3 \leq 1 \, | \, 
%\rho_1+\rho_2+\rho_3 = 1\}$.
Variables $(u_1,u_2)$ span the two-simplex 
$\Sigma_{(2)} = \{ 0 \leq u_2,u_2, u_1+u_2 \leq 1\}$ whereas each variable $u_3$ and $u_4$ spans the interval $\Sigma_{(1)} = [0,1]$ .

\noindent
Again, let us note by $k_1$ and $k_2$ the loop momenta, the $q_i$'s are given by:
\begin{align}
  q_1 &= k_1 - p_1 -p_2, & q_2 &= k_1 - p_2, & q_3 &= k_1, \notag \\
  q_4 &=k_2, & q_5 &= k_2 - p_3, & q_6 &= k_1+k_2+p_4 \notag \\
  q_7 &= k_1 + k_2 & & & &
  \label{eq4ptmom}
\end{align}
To each internal line with four-momentum $q_i$ is associated a mass $m_i$.
Defining $s=(p_1+p_2)^2$, $t=(p_1+p_4)^2$, $u=(p_1+p_3)^2$, the matrix $G$ reads:
%\begin{align}
%G_{11} 
%&= \rho_1^2 \, (\rho_2+\rho_3) \, p_3^2 
%& 
%G_{12} 
%&= \rho_1 \, \rho_2 \, \rho_3 \, (p_3^2 + p_4^2 - s) 
%\notag \\ 
%G_{13} 
%&= \rho_1 \, \rho_3 \, 
%[ (\rho_2+\rho_3) \, (t - p_2^2 - p_3^2) - 2 \, \rho_3 \, p_3^2] 
%& 
%G_{14} 
%&= \rho_1 \, \rho_2 \, \rho_3 \, (p_1^2 + p_3^2 - u) 
%\notag \\ 
%G_{22} 
%&= \rho_2^2 \, (\rho_1+\rho_3) \, p_4^2 
%& 
%G_{23} 
%&= \rho_2 \, \rho_3 \, [ (\rho_1+\rho_3) \, (p_3^2+p_4^2-s) 
%+ \rho_3 \, (u - p_2^2)] 
%\notag \\
%G_{24} 
%&= \rho_1 \, \rho_2 \, \rho_3 \, (t - p_1^2 - p_4^2) 
%& 
%G_{33} 
%&= \rho_3^2 \, 
%[ \rho_3 \, (p_2^2+p_3^2-t) + (\rho_1+\rho_3) \, p_3^2 
%  - (\rho_2 + \rho_3) \, p_2^2] 
%\notag \\ 
%G_{34} 
%&= \rho_3^2 \, 
%[ \rho_2 \, (p_1^2 + p_2^2 - s)+ \rho_1 \, (p_1^2 + p_3^2 - u)] 
%& 
%G_{44} 
%&= \rho_3^2 \, (\rho_1+\rho_2) \, p_1^2
%\label{eq4pt4p}
%\end{align}
\begin{align}
G_{11} 
&= \rho_1^2 \, (\rho_2+\rho_3) \, s
& 
G_{12} 
&= \frac{1}{2} \, \rho_1^2 \, (\rho_2 + \rho_3) \, (s - p_1^2 + p_2^2) 
\notag \\ 
G_{13} 
&= \frac{1}{2} \, \rho_1 \, \rho_2 \, \rho_3 \, (s + p_3^2 - p_4^2)
& 
G_{14} 
&= \frac{1}{2} \, \rho_1 \, \rho_2 \, \rho_3 \, (s - p_3^2 + p_4^2) 
\notag \\ 
G_{22} 
&= \rho_1^2 \, (\rho_2+\rho_3) \, p_2^2 
&
G_{23} 
&= \frac{1}{2} \, \rho_1 \, \rho_2 \, \rho_3 \,(p_2^2 + p_3^2 - t)
\notag \\
G_{24} 
&= \frac{1}{2} \, \rho_1 \, \rho_2 \, \rho_3 \, (p_2^2 + p_4^2 - u) 
&
G_{33} 
&= \rho_2^2 \, (\rho_1+\rho_3) \, p_3^2 
\notag \\ 
G_{34} 
&= \frac{1}{2} \, \rho_1 \, \rho_2 \, \rho_3 \, (p_3^2 + p_4^2 - s)
&
G_{44} 
&= \rho_3^2 \, (\rho_1+\rho_2) \, p_4^2
\label{eq4pt4p}
\end{align}
whereas the vector $V$ reads:
\begin{align}
V_1 
&=  \frac{1}{2} \, \rho_1 \, (\rho_1 \, \rho_2 + \rho_2 \, \rho_3 + \rho_3 \, \rho_1) \, 
\left[ 
s + m_3^2 - m_1^2  
\right] 
\notag \\
V_2 
&=  \frac{1}{2} \, \rho_1 \,(\rho_1 \, \rho_2 + \rho_2 \, \rho_3 + \rho_3 \, \rho_1) \, 
\left[ 
  p_2^2 + m_3^2 - m_2^2
\right] 
\notag \\
V_3 
&=  \frac{1}{2} \, \rho_2 \, (\rho_1 \, \rho_2 + \rho_2 \, \rho_3 + \rho_3 \, \rho_1) \,
\left[ 
  p_3^2 + m_4^2 - m_5^2
\right] 
\notag \\
V_4 
&=  \frac{1}{2} \, \rho_3 \, (\rho_1 \, \rho_2 + \rho_2 \, \rho_3 + \rho_3 \, \rho_1) \,
\left[ 
  p_4^2 + m_7^2 - m_6^2
\right]
  \label{eq4pt5}
\end{align}
and $C$ is given by:
\begin{align}
C 
&= - (\rho_1 \, \rho_2 + \rho_2 \, \rho_3 + \rho_3 \, \rho_1) \, 
     (\rho_1 \, m_3^2 + \rho_2 \, m_4^2 + \rho_3 \, m_7^2)
\label{eq4pt5p}
\end{align}
The explicit expression for $\overline{\calf}$ then follows from eq. 
(\ref{eqG8c}). We thereby get the advocated integral representation 
\begin{align}
^{(2)}I_{4}^{4} \left( \{p_{j}\}; \calt \right)
\quad  
&= 
\int_{(I\!\!R^{+})^{3}} 
d \rho_{1} \, d \rho_{2} \, d \rho_{3} \; 
\rho_1 \, \rho_2 \, \rho_3^{2} \; 
\delta \left( 1 - \sum_{l=1}^{3} \rho_{ \, l} \right) \, 
\left[ \rho_1 \, \rho_2 + \rho_2 \, \rho_3 + \rho_3 \, \rho_1 \right] \;
^{(1)}\widetilde{I}_{5}^{4}
\label{eq4pt6a}
\end{align}
involving the five-point function of ``generalised one-loop type" given by:
\begin{align}
^{(1)}\widetilde{I}_{5}^{4}
& =
\int_{0}^{1} d u_{1} \int_{0}^{1} d u_{2} \int_{\Sigma_{(2)}} du_3 \, du_4
\left[ \,
 \overline{\cal F}(\{u_{k}\},\{\rho_l\}) - i \, \lambda 
\right]^{-3}
\label{eq4pt6b}
\end{align}
The change of variables $\rho_1 = \rho \, \xi$, $\rho_2 = \rho \,(1-\xi)$,
$\rho_3=(1-\rho)$ amounts in eq. (\ref{eq4pt6a}) to the replacement
\[
\int_{(I\!\!R^{+})^{3}} 
d \rho_{1} \, d \rho_{2} \, d \rho_{3} \; 
\rho_1 \, \rho_2 \, \rho_3^{2} \; 
\delta \left( 1 - \sum_{l=1}^{3} \rho_{ \, l} \right) \, 
\left[ \rho_1 \, \rho_2 + \rho_2 \, \rho_3 + \rho_3 \, \rho_1 \right] \times
\; \to \;
\int_{0}^{1} d \rho  \int_{0}^{1} d \xi \, W(\rho,\xi) \times
\]
where the weighting function $W(\rho,\xi)$ is given by:
\begin{align}
W(\rho,\xi) 
&=
\rho^4 \, (1-\rho)^2 \, \xi \, (1-\xi) \, \left[ (1-\rho) + \rho \, \xi \,
(1-\xi) \right]
\label{eq4pt6c}
\end{align}
Here again the four-point function with planar topology with the 
same three-leg type vertices can be worked out in a quite similar way.

\vspace{0.3cm}

\noindent
In the two above examples, the term $\det(A)$ noticeably
factorises in the expressions of the quantities $V$ and $C$. This turns out to
be a general feature at least for the type of three-leg vertices considered. More
precisely, it can be shown that a parametrisation can always be found 
for which this property holds. A general proof is given in Appendix A for any
$N$-point two-loop non planar topology, and a very similar proof holds true also
for any planar topology with this type of vertex. This feature makes the
discussion of both kinematic and fake singularities simpler and more
transparent. In particular the Landau conditions to be fulfilled to encounter 
kinematic singularities take a simple form, whereas this parametrisation gives
handles to circumvent possible numerical instabilities which might be induced by
fake singularities. These issues will be more thoroughly discussed in a future
publication.

\section{Outlook}\label{outlook} 

We do acknowledge that a long way shall still be 
scouted out to extend it to full-fledged two-loop tensor integrals appearing in
general gauge theories with fermions and/or derivative couplings etc. covering
the bestiary of all topologies and cases which appear in the general
two-loop class of interest for precision collider physics. This marathon 
will not be undertaken any further in this account which intends to be a 
first step toward the completion of such a programme. Taking for granted that 
the approach advocated in this article already applies to a collection of 
relevant cases, a further
issue consists in the analytical calculation of the above-coined 
``$N^\prime$-point functions of generalised one-loop type" in closed form. 
As seen in sec.
\ref{argument-general}, the latter are generalised in two respects with respect to
``ordinary'' one-loop functions involved in collider processes at one loop. 
Firstly, the kinematics involved in the computation depends on the extra
integration variables $\rho_j$'s or equivalently $\rho$ and $\xi$ seen here as 
external parameters and may
thus span a wider kinematical phase space than ordinarily met in genuine
one-loop cases. Secondly, the $N^\prime$-point functions of ``generalised 
one-loop-type" are provided by Feynman-type 
parametric integrals over domains differing from the ordinary 
$(N^\prime-1)$-simplex. Although long-tested standard techniques developed 
for the genuine one-loop case might be customised to treat the new ones at hand
as shown in sec. 3, the above two issues motivate the development of a
novel approach which tackles both these issues in a systematic and 
straightforward way while computing the``generalised one-loop type functions". 
The presentation of such an approach and the exploration of its features 
is the subject of the publications \cite{paper1,paper2,paper3}. 
More precisely, the method tailored for the computation of the ``generalised one-loop functions'' is 
implemented for the usual one-loop case with arbitrary kinematics for the real mass case in ref.~\cite{paper1},  
for the case of general complex masses in ref.~\cite{paper2}, and lastly, 
for the IR/collinear divergent cases in ref.~\cite{paper3}.\\

The extension of the strategy advocated above to more general cases will be 
elaborated in subsequent articles
but let us try to discuss it more precisely.
The complexity comes from the growing number of external legs as in the one-loop case but is also related to the number of internal legs, indeed the number of effective external legs for the underlying ``generalised one-loop function'' is the number of internal legs of the two-loop diagram minus 2. So the strategy will be to apply firstly our method to simple topologies, working with fixed number of external legs (two or three) and considering topologies with an increasing number of internal lines. Then, we will move to more complicated cases such as two-loop four-point functions involving ``generalised one-loop functions'' with up to five external legs. All what have been discussed so far is for scalar theories, in real life, we have to face gauge theories which lead to a ``dressing'' of the numerator because of spin 1/2 particles and derivative vertices. The method applied to these cases will involve ``generalised one-loop functions'' with a non-trivial numerator. Although reduction technics in principle apply to ``generalised one-loop functions'' because they are related to the invariance of the one-loop integrals under a shift of the internal 4-momentum running into the loop, their implementations in our case will be postponed after the completion of the scalar cases.
There are ambiguities in our algorithm related to the parametrisation of the internal 4-momenta and also to the choice of the $u$ parameters. It has been shown in appendix~\ref{proof} that for some topologies $\det(A)$ factorises from $V$ and $C$ reducing the complexity of these terms. Although this property does not remove the ambiguities, this is clearly a good guideline to follow. Nevertheless, the choice of a fixed algorithm for all the topologies is premature at this point because the experiences gained with the computation of different topologies will mature it. 
The UV divergent case will not cause problems because our method works also if the space-time dimension is away from 4 as shown in ref.~\cite{paper3}. In addition, by power counting, it is easy to show, for scalar theories having three- and four-leg vertices, that the number of internal lines for the potential UV divergent diagrams is 4 thus the ``generalised one-loop functions'' to be considered are two-point functions.

\section*{In memoriam}
%\label{s5}

This work was initiated by Prof. Shimizu after a visit to LAPTh. He explained to us
his ideas about the numerical computation of scalar two-loop three-and
four-point functions. He shared his notes partly in
English, partly in Japanese with us and he encouraged us to push forward the
project presented  here combining an analytical approach for the ``generalised
one-loop functions" with a numerical computation of the left-over double
parametric integration.  J.-Ph. G. would like to thank Shimizu-sensei for giving
him a taste of Japanese culture and for his kindness.

\section*{Acknowledgements}

We would like to thank K. Kato for fruitful exchanges along this project.

\appendix

\section{Factorisation of $\det(A)$ in $V$ and  $C$\label{proof}}

%\textcolor{blue}{Il me semble que dans une th\'eorie scalaire avec des vertex \`a trois pattes, il n'y a que trois topologies: deux non planaires fig.5 et fig. 6 et une planaire fig.7, est-ce vrai?}\\
Different topologies for a general $N$-point two-loop diagram\footnote{In what follows the 
external momenta $p_j, j=1, \cdots, N$ are assumed to be only constrained by the 
overall energy-momentum conservation $\sum_{j=1}^{N} p_j =0$, otherwise 
arbitrary.} are discussed in the frame of a scalar theory with three-leg vertices. The first topology studied will be treated in details, the main formulae will be set and the reason of this factorisation will be explained. Then, the other topologies will be discussed succinctly.\\

\begin{figure}[h]
\centering
\parbox[c][63mm][t]{63mm}{\begin{fmfgraph*}(80,80)
  \fmfstraight
  \fmfsurroundn{e}{20}
  \fmfset{arrow_len}{3mm}
  \fmf{phantom,tension=16.0}{e8,vv1}
  \fmf{phantom,tension=16.0}{e20,vv3}
  \fmf{fermion,label={\small $p_1$},tension=7.0}{vv1,vv2}
  \fmf{fermion,label={\small $p_{i+1}$},tension=7.0}{vv3,vv4}
  \fmf{phantom,right}{vv1,vv3}
  \fmf{phantom,right}{vv3,vv1}
  \fmf{dots,right}{vv2,vv4}
  \fmf{dots,right}{vv4,vv2}
  \fmfposition
  \fmffreeze
  \fmfipath{p[]}
  \fmfiset{p1}{vpath (__vv2,__vv4)}
  \fmfiset{p2}{vpath (__vv4,__vv2)}
  \fmfipath{pp[]}
  \fmfiset{pp1}{vpath (__vv1,__vv3)}
  \fmfiset{pp2}{vpath (__vv3,__vv1)}
  \fmfipair{a[]}
  \fmfiequ{a2}{(.53w,.95h)}
  \fmfiequ{a3}{(.604w,.47h)}
  \fmfi{fermion,label={\small $q_2$},label.side=left}{subpath (0,length(p1)/6) of p1}
  \fmfi{fermion,label={\small $p_2$},tension=1.0}{point length(pp1)/6 of pp1 -- point length(p1)/6 of p1}
  \fmfi{fermion,label={\small $q_3$},label.side=left}{subpath (length(p1)/6,2length(p1)/6) of p1}
  \fmfi{fermion,label={\small $q_{i}$},label.side=left}{subpath (3length(p1)/6,4length(p1)/6) of p1}
  \fmfi{fermion,label={\small $p_i$},tension=1.0}{point 4length(pp1)/6 of pp1 -- point 4length(p1)/6 of p1}
  \fmfi{fermion,label={\small $q_{i+1}$},label.side=left}{subpath (4length(p1)/6,5length(p1)/6) of p1}
  \fmfi{fermion,label={\small $q_{i+2}$},label.side=right}{subpath (6length(p1)/6,5length(p1)/6) of p1}
  \fmfi{fermion,label={\small $q_{i+3}$},label.side=right}{subpath (length(p2)/6,0) of p2}
  \fmfi{fermion,label={\small $q_{N}$},label.side=right}{subpath (4length(p2)/6,3length(p2)/6) of p2}
  \fmfi{fermion,label={\small $p_{N-1}$},tension=2.0}{point 4length(pp2)/6 of pp2 -- point 4length(p2)/6 of p2}
  \fmfi{fermion,label={\small $q_{N+1}$},label.side=right}{subpath (5length(p2)/6,4length(p2)/6) of p2}
  \fmfiset{p4}{point 5length(pp2)/6 of pp2 -- point 5length(p2)/6 of p2}
  \fmfi{fermion,label={\small $p_{N}$},tension=1.0}{subpath (0,length(p4)/2) of p4}
  \fmfi{fermion,label={\small $q_{N+2}$},label.side=right,tension=1.0}{subpath (length(p4)/2,length(p4)) of p4}
  \fmfi{fermion,label={\small $q_{1}$},label.side=left}{subpath (5length(p2)/6,6length(p2)/6) of p2}
  \fmfiset{p3}{point 5length(p1)/6 of p1 .. a3 ..point 16.7length(p2)/24 of p2 .. a2 .. point length(p4)/2 of p4}
  \fmfi{fermion,rubout,label={\small $q_{N+3}$}}{subpath (0,3length(p3)/4) of p3}
  \fmfi{plain,rubout}{subpath (3length(p3)/4,length(p3)) of p3}
\end{fmfgraph*}}
\caption{\small diagram picturing a two-loop $N$-point function with 
a non planar topology.}
\label{fig4-1} 
\end{figure}
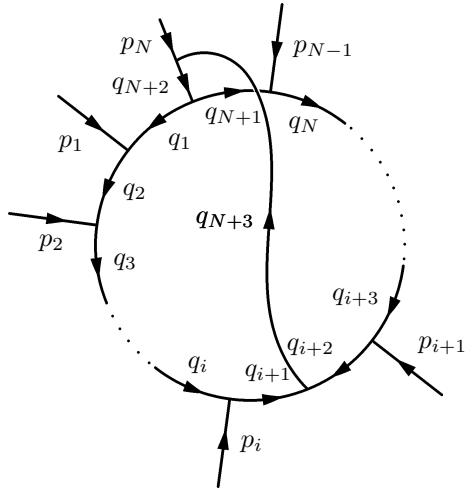

\noindent
%Let us consider a general $N$-point two-loop\footnote{In what follows the 
%external momenta $p_j, j=1, \cdots, N$ are assumed to be only constrained by the 
%overall energy-momentum conservation $\sum_{j=1}^{N} p_j =0$, otherwise 
%arbitrary.} with a non planar topology 
%corresponding to the diagram depicted in fig.~\ref{fig4-1}.
Let us consider first the following diagram with a non planar topology depicted in fig.~\ref{fig4-1}.
\vspace{1cm}
The $n$-momentum carried by each internal line $j=1,\cdots,N+3$ is labelled by 
$q_j$, which can be parametrised in a general way as $q_j = \hat{k}_j + \hr_j$ 
where 
$\hat{r}_j$ is some shift depending on the external momenta $p_l$'s whereas
$\hat{k}_j$ is a linear combination of the loop momenta $k_1$ and $k_2$. 
Let us hereby specify these linear combinations as:
\begin{align}
&\hat{k}_{1} \quad{}\; = \hat{k}_{2} \quad{}\; = \ldots = \hat{k}_{i+1} \;\; \equiv k_1 
\notag \\
&\hat{k}_{i+2} \;\, = \hat{k}_{i+3} \;\,= \ldots = \hat{k}_{N+1} \equiv k_2 
\label{eqfd1}\\
&\hat{k}_{N+2} = \hat{k}_{N+3} = k_1+k_2\notag 
\end{align}
\noindent
The $\hr_j$'s can be conveniently rewritten as:
\begin{align}
\hr_{1} &= \hr_{i+1} - t_1 &\quad \hr_{i+3} &= \hr_{i+2} -t_{i+1} \notag \\
\hr_{2} &= \hr_{i+1} - t_2 &\quad \hr_{i+4} &= \hr_{i+2} -t_{i+2} \notag \\
\hr_{3} &= \hr_{i+1} - t_3 &\quad \hr_{i+5} &= \hr_{i+2} -t_{i+3} \notag \\
\vdots & & \vdots & \notag \\
\hr_{i} &= \hr_{i+1} - t_i &\quad \hr_{N+1} &= \hr_{i+2} -t_{N-1} \notag \\
\hr_{N+2} &= \hr_{i+1} + \hr_{i+2} - (t_1 + t_{N-1}) &\quad \hr_{N+3} &= \hr_{i+1} + \hr_{i+2} 
\label{eqfd2}
\end{align}
with
\begin{align}
  t_{1} &= p_{[1..i]} &\quad t_{i+1} &= p_{i+1} \notag \\
  t_{2} &= p_{[2..i]} &\quad t_{i+2} &= p_{[i+1..i+2]} \notag \\
  t_{3} &= p_{[3..i]} &\quad t_{i+3} &= p_{[i+1..i+3]} \notag \\
  \vdots & & \vdots & \notag \\
  t_{i} &= p_i &\quad t_{N-1} &= p_{[i+1..N-1]}
  \label{eqfd21}
\end{align}
where we have introduced the shorthand:
\begin{align}
  p_{[i..j]} \equiv \sum_{k=i}^{j} p_k 
%& \quad{} \quad{} , \quad{}  \quad{} 
%  \tau_{[i..j]} \equiv \sum_{k=i}^{j} t_k 
%  \label{eqfdnot1}
\end{align}
In eq. (\ref{eqfd2}) $p_N$ has been traded for $- \,(t_1 + t_{N-1})$ using the 
%global 
overall momentum conservation $p_{[1..N]} = 0$. Energy-momentum conservation at
each vertex implies that all the $\hr_j$'s but two can be expressed in terms of
two unfixed ones. These two arbitrary $\hr$'s, which we implicitly
chose\footnote{This choice is not unique: any other choice of arbitrary duo
$\hr_j,\hr_k$ associated to distinct loops e.g. with $j \in \{1,\cdots,i+1\}$
and $k \in \{i+2,\cdots,N+1\}$ would also fit the purpose.} above to be 
$\hr_{i+1}$ and $\hr_{i+2}$, reflect nothing but the invariance of the Feynman
diagram under two independent shifts of the loop momenta $k_1$ and $k_2$ by
arbitrary constants. The latter may be parametrised in a general way as:
\begin{align}
  \hr_{i+1} &= \sum_{i=1}^{N-1} \alpha_i \, t_i, \quad
  \hr_{i+2} = \sum_{i=1}^{N-1} \beta_i \, t_i
  \label{eqfd6}
\end{align}
This notational preamble being set, let us consider the quantity:
\begin{align}
\sum_{j=1}^{N+3} \tau_j \, (q_j - m_j^2) 
& = \sum_{j=1}^{N+3} \tau_j \, 
   [ \hat{k}_j^2 + 2 \, (\hat{k}_j \cdot \hat{r}_j)  + \hat{r}_j^2 - m_j^2]
  \label{eqfd3}
\end{align}
Using specification (\ref{eqfd1}) we write
%$\sum_{i=1}^{N+3} \tau_i \, \hat{k}_i^2$ as 
\[
 \sum_{j=1}^{N+3} \tau_j \, \hat{k}_j^2
 \equiv
 [k_1 \quad  k_2] \cdot A \cdot \left[
  \begin{array}{c}
    k_1 \\
    k_2
  \end{array}
\right]
\]
This defines the $2 \times 2$ matrix $A$ whose elements are
\begin{align}
A_{11} 
&=  \rho_1 + \rho_3,
\quad
A_{22} = \rho_2 + \rho_3
\quad
\text{and} 
\quad
A_{12} = \rho_3
\label{eqfd4}
\end{align}
where we introduced the three parameters $\rho_k, k=1,2,3$ defined by:
\begin{align}
\rho_k 
&= \sum_{j \in S_k} \tau_{j}, \quad k=1,2,3
%\quad 
%\rho_2 = \sum_{j \in S_2} \tau_{j} 
%\quad 
%\text{and} 
%\quad 
%\rho_3 = \sum_{j \in S_3} \tau_{j}
\label{eqfd5}\\
S_1 
&= \{1, \cdots,i+1\}, 
\quad
S_2 = \{i+2, \cdots,N+1\}, 
\quad
\text{and} 
\quad 
S_3 = \{N+2,N+3\} 
\label{eqfd5bis}
\end{align}
closely following the discussion in sec. \ref{argument-general}.

\vspace{0.3cm}

\noindent
The second term of the r.h.s. of eq. (\ref{eqfd3}) may be recast:
\begin{align}
\sum_{j=1}^{N+3} \tau_i \, (\hat{k}_j \cdot \hat{r}_j)
\equiv  
[k_1 \quad  k_2] \cdot B 
\cdot 
\left[
  \begin{array}{c}
    t_1 \\
    t_2 \\
    \vdots \\
    t_{N-1}
  \end{array}
\right]
\label{intro-T}
\end{align}
where $B$ is a $2 \times (N-1)$ matrix whose Feynman parameter dependent 
elements are read using eqs. (\ref{eqfd2}) and parametrisation (\ref{eqfd6}):
\begin{align}
  B_{1k} &= \left\{
  \begin{array}{rcl}
    -(\tau_1 + \tau_{N+2}) + \alpha_k \, (\rho_1 + \rho_3) + \beta_k \, \rho_3 & \text{if} & k =1 \\
    -\tau_{k}  + \alpha_k \, (\rho_1 + \rho_3) + \beta_k \, \rho_3 & \text{if} & 1 < k \leq i \\
     \alpha_k \, (\rho_1 + \rho_3) + \beta_k \, \rho_3 & \text{if} & i < k < N-1 \\
    -\tau_{N+2} + \alpha_k \, (\rho_1 + \rho_3) + \beta_k \, \rho_3 & \text{if} & k = N-1 \\
  \end{array}
  \right. \notag \\
  B_{2k} &= \left\{
  \begin{array}{rcl}
    -\tau_{N+2} + \alpha_k \, \rho_3 + \beta_k \, (\rho_2 + \rho_3) & \text{if} & k =1 \\
    \alpha_k \, \rho_3 + \beta_k \, (\rho_2 + \rho_3) & \text{if} & 1 < k \leq i \\
    -\tau_{k+2} + \alpha_k \, \rho_3 + \beta_k \, (\rho_2 + \rho_3) & \text{if} & i < k < N-1 \\
    -(\tau_{N+2} + \tau_{N+1}) + \alpha_k \, \rho_3 + \beta_k \, (\rho_2 + \rho_3) & \text{if} & k = N-1 \\
  \end{array}
  \right.
  \label{eqfd7}
\end{align}
For a general parametrisation of the Feynman diagram, $\alpha_j$
and $\beta_j$ are not all vanishing, the $B$ matrix thus depends on all the 
Feynman parameters $\tau_j$. We can clarify this Feynman parameter dependence
noting that the matrix $B$ can be recast as follows:
\begin{align}
  B &= \overline{B} + A \cdot \Delta
  \label{eqfd8}
\end{align}
where $\overline{B}$ and $\Delta$ are defined by
\begin{align}
  \Delta &=  \left[
  \begin{array}{ccccc}
    \alpha_1 & \alpha_2 & \ldots & \alpha_{N-2} & \alpha_{N-1} \\
    \beta_1 & \beta_2 & \ldots & \beta_{N-2} & \beta_{N-1}
  \end{array}
\right]
  \label{eqfd81}
\end{align}
and
\begin{align}
  \overline{B}_{1k} &= 
  \left\{
  \begin{array}{ccc}
    -(\tau_1 + \tau_{N+2}) & \text{if} & k =1 \\
    -\tau_{k}  & \text{if} & 1 < k \leq i \\
     0 & \text{if} & i < k < N-1 \\
    -\tau_{N+2} & \text{if} & k = N-1 \\
  \end{array}
  \right. \notag \\
  \overline{B}_{2k} &= 
  \left\{
  \begin{array}{ccc}
    -\tau_{N+2} & \text{if} & k =1 \\
    0 & \text{if} & 1 < k \leq i \\
    -\tau_{k+2} & \text{if} & i < k < N-1 \\
    -(\tau_{N+2} + \tau_{N+1}) & \text{if} & k = N-1 \\
  \end{array}
  \right.
  \label{eqfd9}
\end{align}
%\begin{align}
  %\overline{B}_{1k} &= \left\{
  %\begin{array}{ccc}
    %-( \tau_{N+2} + \tau_{[1..k]})  & \text{if} & k \leq i \\
    %- \tau_{N+2}  & \text{if} & k > i \\
  %\end{array}
  %\right. \notag \\
  %\overline{B}_{2k} &= \left\{
  %\begin{array}{ccc}
    %-\tau_{N+2}  & \text{if} & k \leq i \\
    %- (\tau_{N+2} + \tau_{[k+2..N+1]})  & \text{if} & k > i \\
  %\end{array}
  %\right.
  %\label{eqfd9}
%\end{align}
%\begin{align}
  %B_0 &= -\left[
  %\begin{array}{ccccccc}
    %\tau_{N+2} + \tau_{N+1} & \tau_{N+2} + \tau_{N+1} + \tau_1 & \ldots & \tau_{N+2} + \tau_{N+1} + \tau_{[1..i-1]} &  \tau_{N+2} & \ldots & \tau_{N+2} \\
    %\tau_{N+2}  & \tau_{N+2}  & \ldots & \tau_{N+2}  &  \tau_{N+2} + \tau{[i+2..N]} & \ldots & \tau_{N+2} + \tau_{N}
  %\end{array}
%\right]
  %\label{eqfd9}
%\end{align}
The important property of $\overline{B}$ is that it depends neither on 
$\tau_{i+1}$ 
nor on $\tau_{i+2}$ nor on $\tau_{N+3}$. Let us note $T$ the $(N-1)$ 
column-vector whose elements are $t_1, \cdots, t_{N-1}$, which entered 
eq. (\ref{intro-T}).
%From the structure of ${\cal F}$ given in eq.(\ref{eqG2bis}), the first term 
%between brakets 
The bracketed term in the r.h.s. of eq. (\ref{eqG2bis}) can be re-written as:
\begin{align}
  [r_1 \quad  r_2] \cdot \mbox{Cof}[A] \cdot \left[
  \begin{array}{c}
    r_1 \\
    r_2
  \end{array}
  \right]
  &= (B \cdot T)^T \cdot \mbox{Cof}[A] \cdot (B \cdot T) \label{eqfd10} \\
  &= (\overline{B} \cdot T)^T \cdot \mbox{Cof}[A] \cdot (\overline{B} \cdot T) 
%\notag \\
%  &\quad 
  + \det(A) \, T^T \cdot \left[2 \, \Delta^T \cdot \overline{B} +
  \Delta^T \cdot A \cdot \Delta \right] \cdot T
  \notag
  %\label{eqfd10}
\end{align}
The term $\calc$ in eq. (\ref{eqG2bis}) can in its turn be written using 
eqs. (\ref{eqfd2}) and (\ref{eqfd6}) as
\begin{align}
\calc 
&= 
[\hr_{i+1} \quad  \hr_{i+2}] 
\cdot 
A 
\cdot 
\left[
  \begin{array}{c}
    \hr_{i+1} \\
    \hr_{i+2}
  \end{array}
\right]
+ 2 \, 
[\hr_{i+1} \quad  \hr_{i+2}] \cdot (\overline{B} \cdot T)
+ 
T^T \cdot \Gamma \cdot T - \sum_{j=1}^{I} \tau_j \, m_j^2
\notag \\
&= T^T \cdot [ \Delta^T \cdot A \cdot \Delta + 2 \, \Delta^T \cdot \overline{B} + \Gamma] \cdot T - \sum_{j=1}^{I} \tau_j \, m_j^2 
  \label{eqfd11}
\end{align}
where $\Gamma$ is a $(N-1) \times (N-1)$ matrix read on eq. (\ref{eqfd2})
%, depending on the Feynman parameters, 
and given by
\begin{align}
  \Gamma &= \left[
  \begin{array}{cccccccc}
    \tau_1 + \tau_{N+2} & 0 & \cdots & 0 & 0 & \cdots & 0 & \tau_{N+2} \\
    0 & \tau_2 & \cdots & 0 & 0 & \cdots & 0 & 0 \\
    \vdots & \vdots & \ddots & \vdots & \vdots & \cdots & \vdots & 0 \\    
    0 & 0 & \cdots & \tau_{i} & 0 & \cdots & 0 & 0 \\
    0 & 0 & \cdots & 0 & \tau_{i+3} & \cdots & 0 & 0 \\
    \vdots & \vdots & \vdots & \vdots & 0 & \ddots & \vdots & \vdots \\
    0 & \vdots & \vdots & \vdots & \vdots & \cdots & \tau_N & 0 \\
    \tau_{N+2} & 0 & \cdots & 0 & 0 & \cdots & 0 & \tau_{N+1} + \tau_{N+2}
  \end{array}
\right]
  \label{eqfd111}
\end{align}
As $\overline{B}$, the matrix $\Gamma$ depends neither on $\tau_{i+1}$ nor on 
$\tau_{i+2}$ nor $\tau_{N+3}$.
%, as read on eq. (\ref{eqfd111}). 
The $\Delta$-dependent terms in eqs. (\ref{eqfd10}) and (\ref{eqfd11}) 
cancel each other in $\calf$, reflecting the independence of the Feynman 
diagram considered under the arbitrary shifts on loop momenta parametrised by
$\hr_{i+1},\hr_{i+2}$. The quantity $\calf$ simplifies into:
\begin{align}
\calf &= (\overline{B} \cdot T)^T \cdot \mbox{Cof}[A] \cdot (\overline{B} \cdot T)
- 
\det(A) \, 
\left[ T^T  \cdot \Gamma  \cdot T  -  \sum_{i=1}^{I} \tau_i \, m_i^2 \right]
\label{eqfd12}
\end{align}
Here comes the key point. 
Since none of the three parameters $\tau_{i+1}$, $\tau_{i+2}$ and $\tau_{N+3}$ 
enters into the matrices $\overline{B}$ and $\Gamma$, the $\tau_j$'s may be 
conveniently reparametrised in three subsets corresponding respectively to 
$j \in S_k, k=1,2,3$ defined in eq. (\ref{eqfd5bis}),
introducing $u_l$'s parameters such that
the $\overline{B}$ matrix be homogeneous of degree 1 in the $u_l$'s,
namely\footnote{Had we chosen $\hr_m, \hr_n$ with any $m \in S_1, n \in S_2$ to 
be arbitrary instead of $\hr_{i+1}, \hr_{i+2}$ cf. footnote {\tiny 8}, a similar
reparametrisation could be obtained leading to similar features, {\it mutatis
mutandis}.}:
\begin{align}
\mbox{}
& 
\left.
 \begin{array}{lcl}
  \tau_k &=& \rho_1 \, u_k \quad \text{for $k \in \{1,\ldots,i\}$} 
%\notag
  \\
  \tau_{i+1} &=& \rho_1 \, (1 - \sum_{j=1}^{i} u_j) 
%\notag 
  \\
  && 0 \leq u_k, \; \sum_{j=1}^{i} u_j \leq 1, 
\quad \text{$k \in \{1,\ldots,i\}$}
 \end{array}
\right\}
%\notag 
%\\
\label{eqfd13a}
\end{align}
\begin{align}
\mbox{}
& 
\left.
 \begin{array}{lcl}
  \tau_k &=& \rho_2 \, u_{k-2} \quad \text{for $k \in \{i+3,\ldots,N+1\}$} 
%\notag
  \\
  \tau_{i+2} &=& \rho_2 \, (1 - \sum_{j=i+1}^{N-1} u_j) 
%\notag 
  \\
  && 0 \leq u_k, \; \sum_{j=1}^{i} u_j \leq 1, 
\quad \text{$k \in \{i+3,\ldots,N+1\}$}
 \end{array}
\right\}
%\notag 
%\\
\label{eqfd13b}
\end{align}
\begin{align}
\mbox{}
& 
\left.
 \begin{array}{lcl}
  \tau_{N+2} &=& \rho_3 \, u_N 
%\notag
  \\
  \tau_{N+3} &=& \rho_3 \, (1 - u_N) 
%\notag
  \\
  && 0 \leq u_N \leq 1
 \end{array}
\right\}
\label{eqfd13c}
\end{align}
With this choice, the term $\overline{B}^T \cdot \mbox{Cof}[A] \cdot \overline{B}$ is
homogeneous of degree 2 in the $u_i$'s and the corresponding term 
$(\overline{B} \cdot T)^T \cdot \mbox{Cof}[A] \cdot (\overline{B} \cdot T)$ in eq. 
(\ref{eqfd12}) thus contributes only to the 
$U^T \cdot G \cdot U$ of eq. (\ref{eqG8c}), whereas the other term 
in eq. (\ref{eqfd12}) only contributes to the terms $V$ and $C$ in eq. 
(\ref{eqG8c}), which both appear to be $\propto \det(A)$. Furthermore the term 
$T^T \cdot \Gamma \cdot T$ being homogeneous of degree 1 in the $u_i$'s 
contributes only to $V$ but not to $C$: the $C$ term is thus a mere linear 
combination of $m_{i+1}^2$, $m_{i+2}^2$ and $m_{N+3}^2$.

\vspace{0.3cm}

\noindent
Using these results we can proceed further and compute the matrix
$G$, the vector $V$ and the scalar $C$ 
%as 
defined in eq. (\ref{eqG8c}).
% using the preceding results. 
The latter are the algebraic ingredients in terms 
of which the novel approach advocated in the outlook and presented in 
\cite{paper1,paper2,paper3} naturally proceeds.
This will be the purpose of future publications. \\

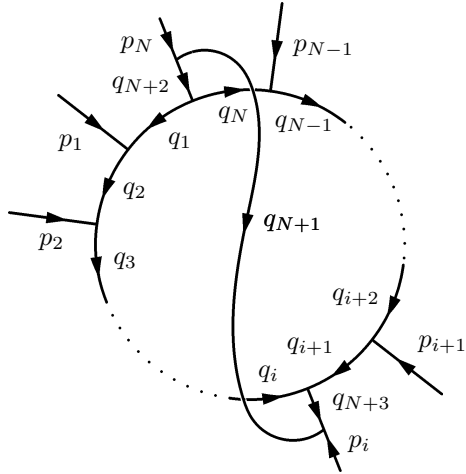
\begin{figure}[h]
\centering
\parbox[c][63mm][t]{63mm}{\begin{fmfgraph*}(80,80)
  \fmfstraight
  \fmfsurroundn{e}{20}
  \fmfset{arrow_len}{3mm}
  \fmf{phantom,tension=16.0}{e8,vv1}
  \fmf{phantom,tension=16.0}{e20,vv3}
  \fmf{fermion,label={\small $p_1$},tension=7.0}{vv1,vv2}
  \fmf{fermion,label={\small $p_{i+1}$},tension=7.0}{vv3,vv4}
  \fmf{phantom,right}{vv1,vv3}
  \fmf{phantom,right}{vv3,vv1}
  \fmf{dots,right}{vv2,vv4}
  \fmf{dots,right}{vv4,vv2}
  \fmfposition
  \fmffreeze
  \fmfipath{p[]}
  \fmfiset{p1}{vpath (__vv2,__vv4)}
  \fmfiset{p2}{vpath (__vv4,__vv2)}
  \fmfipath{pp[]}
  \fmfiset{pp1}{vpath (__vv1,__vv3)}
  \fmfiset{pp2}{vpath (__vv3,__vv1)}
  \fmfipair{a[]}
  \fmfiequ{a2}{(.53w,.95h)}
  \fmfiequ{a3}{(.544w,.69h)}
  \fmfiequ{a4}{(.554w,.36h)}
  \fmfi{fermion,label={\small $q_2$},label.side=left}{subpath (0,length(p1)/6) of p1}
  \fmfi{fermion,label={\small $p_2$},tension=1.0}{point length(pp1)/6 of pp1 -- point length(p1)/6 of p1}
  \fmfi{fermion,label={\small $q_3$},label.side=left}{subpath (length(p1)/6,2length(p1)/6) of p1}
  \fmfi{fermion,label={\small $q_{i}$},label.side=left}{subpath (4length(p1)/6,5length(p1)/6) of p1}
  \fmfi{fermion,label={\small $q_{i+1}$},label.side=right}{subpath (6length(p1)/6,5length(p1)/6) of p1}
  \fmfi{fermion,label={\small $q_{i+2}$},label.side=right}{subpath (length(p2)/6,0) of p2}
  \fmfi{fermion,label={\small $q_{N-1}$},label.side=right}{subpath (4length(p2)/6,3length(p2)/6) of p2}
  \fmfi{fermion,label={\small $p_{N-1}$},tension=2.0}{point 4length(pp2)/6 of pp2 -- point 4length(p2)/6 of p2}
  \fmfi{fermion,label={\small $q_{N}$},label.side=right}{subpath (5length(p2)/6,4length(p2)/6) of p2}
  \fmfiset{p4}{point 5length(pp2)/6 of pp2 -- point 5length(p2)/6 of p2}
  \fmfi{fermion,label={\small $p_{N}$},tension=1.0}{subpath (0,length(p4)/2) of p4}
  \fmfi{fermion,label={\small $q_{N+2}$},label.side=right,tension=1.0}{subpath (length(p4)/2,length(p4)) of p4}
  \fmfi{fermion,label={\small $q_{1}$},label.side=left}{subpath (5length(p2)/6,6length(p2)/6) of p2}
  \fmfiset{p5}{point 5length(pp1)/6 of pp1 -- point 5length(p1)/6 of p1}
\fmfi{fermion,label={\small $p_i$},tension=1.0}{subpath (0,length(p5)/2) of p5}
  \fmfi{fermion,label={\small $q_{N+3}$},label.side=left,tension=1.0}{subpath (length(p5),length(p5)/2) of p5}
  \fmfiset{p3}{point length(p5)/2 of p5 .. a4 .. point 16.7length(p1)/24 of p1 .. a3 .. point 16.9length(p2)/24 of p2 .. a2 .. point length(p4)/2 of p4}
  %\fmfi{fermion,rubout,label={\small $q_{N+1}$},label.side=left}{subpath (4.8length(p3)/5,0) of p3}
  \fmfi{plain}{subpath (0,length(p3)/100) of p3}
  \fmfi{fermion,rubout,label={\small $q_{N+1}$},label.side=left}{subpath (4.8length(p3)/5,length(p3)/100) of p3}
  \fmfi{plain,rubout}{subpath (4.8length(p3)/5,99length(p3)/100) of p3}
  \fmfi{plain}{subpath (99length(p3)/100,length(p3)) of p3}
\end{fmfgraph*}}
\caption{\small diagram picturing a two-loop $N$-point function with 
a non planar topology.}
\label{fig6} 
\end{figure}

\noindent
Let us move now to the other topologies. For those topologies, once the parametrisation of the internal lines has been chosen, the sequences for the proof of the factorisation are identical to the previous case, so they will not be reproduced here and only the matrices $\overline{B}$ will be given.
%
%\noindent
The other non planar topology\footnote{Strictly speaking, this kind of topology can be planar in specific cases: for example if $i$ is chosen to be $N-1$ in fig.~\ref{fig6}, or in other words if the internal leg labelled by $N+1$ connects two adjacent external legs. Note that this is always the case if $N=3$.} corresponds to the diagram depicted in fig.~\ref{fig6}. 
\vspace{0.5cm}
Using the same parametrisation for the $q_i$'s as in the preceding topology studied, the $\hat{k}_{i}$'s are given by:
\begin{align}
&\hat{k}_{1} \quad{}\; = \hat{k}_{2} \quad{}\; = \ldots = \hat{k}_{i} \;\; \equiv k_1 
\notag \\
&\hat{k}_{i+1} \;\, = \hat{k}_{i+2} \;\,= \ldots = \hat{k}_{N} \equiv k_2 
\label{eqfda1}\\
&\hat{k}_{N+1} = \hat{k}_{N+2} = \hat{k}_{N+3} = k_1+k_2\notag 
\end{align}
and the $\hr_i$'s by:
\begin{align}
\hr_{1} &= \hr_{i} - t_1 &\quad \hr_{i+2} &= \hr_{i+1} -t_{i} \notag \\
\hr_{2} &= \hr_{i} - t_2 &\quad \hr_{i+3} &= \hr_{i+1} -t_{i+1} \notag \\
\hr_{3} &= \hr_{i} - t_3 &\quad \hr_{i+4} &= \hr_{i+1} -t_{i+2} \notag \\
\vdots & & \vdots & \notag \\
\hr_{i-1} &= \hr_{i} - t_{i-1} &\quad \hr_{N} &= \hr_{i+1} -t_{N-2} \notag \\
\hr_{N+1} &= \hr_{i} + \hr_{i+1} - t_{N-1} &\quad \hr_{N+2} &= \hr_{i} + \hr_{i+1} - (t_1 + t_{N-2})  \notag \\
\hr_{N+3} &= \hr_{i} + \hr_{i+1} & 
\label{eqfda2}
\end{align}
with
\begin{align}
  t_{1} &= p_{[1..i-1]} &\quad t_{i} &= p_{i+1} \notag \\
  t_{2} &= p_{[2..i-1]} &\quad t_{i+1} &= p_{[i+1..i+2]} \notag \\
  t_{3} &= p_{[3..i-1]} &\quad t_{i+2} &= p_{[i+1..i+3]} \notag \\
  \vdots & & \vdots & \notag \\
  t_{i-1} &= p_{i-1} &\quad t_{N-2} &= p_{[i+1..N-1]} \notag \\
  t_{N-1} &= p_i & 
  \label{eqfda21}
\end{align}
The three sets of internal line labels become in this case:
\begin{align}
  S_1 &= \{1, \cdots,i\}, \quad S_2 = \{i+1, \cdots, N\} \quad \text{and} \quad S_3 = \{N+1, N+2, N+3 \} 
  \label{eqdefseta}
\end{align}
The 4-momenta $\hr_{i}$ and $\hr_{i+1}$ can be, in turn, parametrised with the $t_i$, $i=1 \ldots N-1$ as in eq.~\ref{eqfd6}.
It is already clear from this point that, since from energy-momentum conservation 
we achieved to have one internal line for each set $S_i$, $i=1,2,3$ whose $\hr$ 
does not depend on the $t_i$'s. Thus, because of its definition, the matrix 
$\overline{B}$ will not depend on the Feynman parameter associated to these 
three internal lines. Indeed, a simple computation along the the same lines as 
the first example treated shows that:
\begin{align}
  \overline{B}_{1k} &= 
  \left\{
  \begin{array}{ccc}
    -(\tau_1 + \tau_{N+2}) & \text{if} & k =1 \\
    -\tau_{k}  & \text{if} & 1 < k \leq i-1 \\
     0 & \text{if} & i \leq k < N-2 \\
    -\tau_{N+2} & \text{if} & k = N-2 \\
    -\tau_{N+1} & \text{if} & k = N-1 \\
  \end{array}
  \right. \notag \\
  \overline{B}_{2k} &= 
  \left\{
  \begin{array}{ccc}
    -\tau_{N+2} & \text{if} & k =1 \\
    0 & \text{if} & 1 < k < i \\
    -\tau_{k+2} & \text{if} & i \leq k < N-2 \\
    -(\tau_{N+2} + \tau_{N}) & \text{if} & k = N-2 \\
    -\tau_{N+1} & \text{if} & k = N-1 \\
  \end{array}
  \right.
  \label{eqfda9}
\end{align}
which confirms that the three Feynman parameters $\tau_i$, $\tau_{i+1}$ and $\tau_{N+3}$ do not appear in the $\overline{B}$ matrix.
Since these three Feynman parameters belong to the different sets $S_i$, $i=1,2,3$, it is always possible to choose the parameters 
$u_i$' s in such a way that $\overline{B}^T \cdot \mbox{Cof}[A] \cdot \overline{B}$ is homogeneous of degree 2 in these variables. 
Note that, in this case too, the $\Gamma$ matrix does not depend on $\tau_i$, $\tau_{i+1}$ and $\tau_{N+3}$.\\

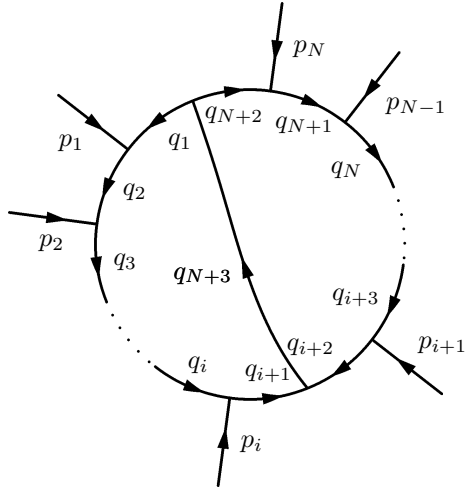
\begin{figure}[h]
\centering
\parbox[c][63mm][t]{63mm}{\begin{fmfgraph*}(80,80)
  \fmfstraight
  \fmfsurroundn{e}{20}
  \fmfset{arrow_len}{3mm}
  \fmf{phantom,tension=16.0}{e8,vv1}
  \fmf{phantom,tension=16.0}{e20,vv3}
  \fmf{fermion,label={\small $p_1$},tension=7.0}{vv1,vv2}
  \fmf{fermion,label={\small $p_{i+1}$},tension=7.0}{vv3,vv4}
  \fmf{phantom,right}{vv1,vv3}
  \fmf{phantom,right}{vv3,vv1}
  \fmf{dots,right}{vv2,vv4}
  \fmf{dots,right}{vv4,vv2}
  \fmfposition
  \fmffreeze
  \fmfipath{p[]}
  \fmfiset{p1}{vpath (__vv2,__vv4)}
  \fmfiset{p2}{vpath (__vv4,__vv2)}
  \fmfipath{pp[]}
  \fmfiset{pp1}{vpath (__vv1,__vv3)}
  \fmfiset{pp2}{vpath (__vv3,__vv1)}
  \fmfipair{a[]}
  \fmfiequ{a2}{(.50w,.75h)}
  \fmfiequ{a3}{(.604w,.47h)}
  \fmfi{fermion,label={\small $q_2$},label.side=left}{subpath (0,length(p1)/6) of p1}
  \fmfi{fermion,label={\small $p_2$},tension=1.0}{point length(pp1)/6 of pp1 -- point length(p1)/6 of p1}
  \fmfi{fermion,label={\small $q_3$},label.side=left}{subpath (length(p1)/6,2length(p1)/6) of p1}
  \fmfi{fermion,label={\small $q_{i}$},label.side=left}{subpath (3length(p1)/6,4length(p1)/6) of p1}
  \fmfi{fermion,label={\small $p_i$},tension=1.0}{point 4length(pp1)/6 of pp1 -- point 4length(p1)/6 of p1}
  \fmfi{fermion,label={\small $q_{i+1}$},label.side=left}{subpath (4length(p1)/6,5length(p1)/6) of p1}
  \fmfi{fermion,label={\small $q_{i+2}$},label.side=right}{subpath (6length(p1)/6,5length(p1)/6) of p1}
  \fmfi{fermion,label={\small $q_{i+3}$},label.side=right}{subpath (length(p2)/6,0) of p2}
  \fmfi{fermion,label={\small $q_{N}$},label.side=right}{subpath (3length(p2)/6,2length(p2)/6) of p2}
  \fmfi{fermion,label={\small $q_{N+1}$},label.side=right}{subpath (4length(p2)/6,3length(p2)/6) of p2}
  \fmfi{fermion,label={\small $p_{N-1}$},tension=2.0}{point 3length(pp2)/6 of pp2 -- point 3length(p2)/6 of p2}
  \fmfi{fermion,label={\small $p_{N}$},tension=2.0}{point 4length(pp2)/6 of pp2 -- point 4length(p2)/6 of p2}
  \fmfi{fermion,label={\small $q_{N+2}$},label.side=right}{subpath (5length(p2)/6,4length(p2)/6) of p2}
  \fmfiset{p4}{point 5length(pp2)/6 of pp2 -- point 5length(p2)/6 of p2}
  \fmfi{fermion,label={\small $q_{1}$},label.side=left}{subpath (5length(p2)/6,6length(p2)/6) of p2}
  \fmfiset{p3}{point 5length(p1)/6 of p1 .. a3 .. a2 .. point length(p4) of p4}
  \fmfi{fermion,rubout,label={\small $q_{N+3}$}}{subpath (0,3length(p3)/4) of p3}
  \fmfi{plain,rubout}{subpath (3length(p3)/4,length(p3)) of p3}
\end{fmfgraph*}}
\caption{\small diagram picturing a two-loop $N$-point function with 
a planar topology.}
\label{fig7} 
\end{figure}

\noindent
For the third topology depicted in fig.~\ref{fig7},
\vspace{1cm}
taking the same parametrisation for the $q_i$'s as in the first case studied, these linear combinations are specified as:
\begin{align}
&\hat{k}_{1} \quad{}\; = \hat{k}_{2} \quad{}\; = \ldots = \hat{k}_{i+1} \;\; \equiv k_1 
\notag \\
&\hat{k}_{i+2} \;\, = \hat{k}_{i+3} \;\,= \ldots = \hat{k}_{N+2} \equiv k_2 
\label{eqtp1}\\
&\hat{k}_{N+3} = k_1+k_2\notag 
\end{align}
\noindent
The $\hr_j$'s can be conveniently rewritten as:
\begin{align}
\hr_{1} &= \hr_{i+1} - t_1 &\quad \hr_{i+3} &= \hr_{i+2} -t_{i+1} \notag \\
\hr_{2} &= \hr_{i+1} - t_2 &\quad \hr_{i+4} &= \hr_{i+2} -t_{i+2} \notag \\
\hr_{3} &= \hr_{i+1} - t_3 &\quad \hr_{i+5} &= \hr_{i+2} -t_{i+3} \notag \\
\vdots & & \vdots & \notag \\
\hr_{i} &= \hr_{i+1} - t_i &\quad \hr_{N+1} &= \hr_{i+2} -t_{N-1} \notag \\
\hr_{N+2} &= \hr_{i+2} + t_{1} &\quad \hr_{N+3} &= \hr_{i+1} + \hr_{i+2} 
\label{eqtp2}
\end{align}
with
\begin{align}
  t_{1} &= p_{[1..i]} &\quad t_{i+1} &= p_{i+1} \notag \\
  t_{2} &= p_{[2..i]} &\quad t_{i+2} &= p_{[i+1..i+2]} \notag \\
  t_{3} &= p_{[3..i]} &\quad t_{i+3} &= p_{[i+1..i+3]} \notag \\
  \vdots & & \vdots & \notag \\
  t_{i} &= p_i &\quad t_{N-1} &= p_{[i+1..N-1]}
  \label{eqtp21}
\end{align}
The sets of the internal line labels are given in this case by:
\begin{align}
S_1 
&= \{1, \cdots,i+1\}, 
\quad
S_2 = \{i+2, \cdots,N+2\}, 
\quad
\text{and} 
\quad 
S_3 = \{N+3\} 
  \label{eqdefsetb}
\end{align}
Again, for each set $S_i$, $i= 1,2,3$, one internal line 4-momentum is independent on the $t_i$'s and so the property of factorisation of $\det(A)$ holds also in this case as can be proven by an explicit computation:
\begin{align}
  \overline{B}_{1k} &= 
  \left\{
  \begin{array}{ccc}
    -\tau_{k}  & \text{if} & 1 \leq k \leq i \\
     0 & \text{if} & i < k \leq N-1 \\
  \end{array}
  \right. \notag \\
  \overline{B}_{2k} &= 
  \left\{
  \begin{array}{ccc}
    -\tau_{N+2} & \text{if} & k =1 \\
    0 & \text{if} & 1 < k \leq i \\
    -\tau_{k+2} & \text{if} & i < k \leq N-1 \\
  \end{array}
  \right.
  \label{eqtp9}
\end{align}
showing explicitly that $\overline{B}$ (and $\Gamma$) do not depend neither on 
$\tau_{i+1}$ 
nor on $\tau_{i+2}$ nor on $\tau_{N+3}$.\\

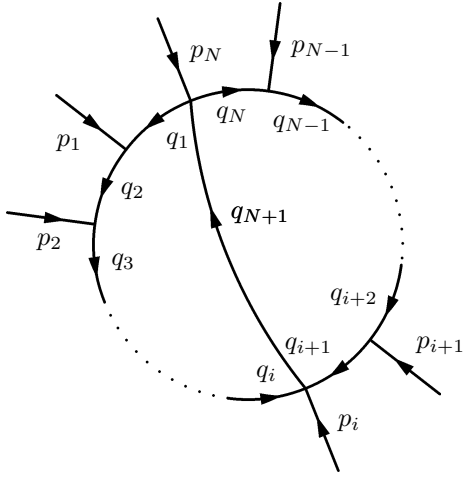
\begin{figure}[h]
\centering
\parbox[c][63mm][t]{63mm}{\begin{fmfgraph*}(80,80)
  \fmfstraight
  \fmfsurroundn{e}{20}
  \fmfset{arrow_len}{3mm}
  \fmf{phantom,tension=16.0}{e8,vv1}
  \fmf{phantom,tension=16.0}{e20,vv3}
  \fmf{fermion,label={\small $p_1$},tension=7.0}{vv1,vv2}
  \fmf{fermion,label={\small $p_{i+1}$},tension=7.0}{vv3,vv4}
  \fmf{phantom,right}{vv1,vv3}
  \fmf{phantom,right}{vv3,vv1}
  \fmf{dots,right}{vv2,vv4}
  \fmf{dots,right}{vv4,vv2}
  \fmfposition
  \fmffreeze
  \fmfipath{p[]}
  \fmfiset{p1}{vpath (__vv2,__vv4)}
  \fmfiset{p2}{vpath (__vv4,__vv2)}
  \fmfipath{pp[]}
  \fmfiset{pp1}{vpath (__vv1,__vv3)}
  \fmfiset{pp2}{vpath (__vv3,__vv1)}
  \fmfipair{a[]}
  %\fmfiequ{a2}{(.53w,.95h)}
  \fmfiequ{a3}{(.494w,.69h)}
  %\fmfiequ{a4}{(.554w,.36h)}
  \fmfi{fermion,label={\small $q_2$},label.side=left}{subpath (0,length(p1)/6) of p1}
  \fmfi{fermion,label={\small $p_2$},tension=1.0}{point length(pp1)/6 of pp1 -- point length(p1)/6 of p1}
  \fmfi{fermion,label={\small $q_3$},label.side=left}{subpath (length(p1)/6,2length(p1)/6) of p1}
  \fmfi{fermion,label={\small $q_{i}$},label.side=left}{subpath (4length(p1)/6,5length(p1)/6) of p1}
  \fmfi{fermion,label={\small $q_{i+1}$},label.side=right}{subpath (6length(p1)/6,5length(p1)/6) of p1}
  \fmfi{fermion,label={\small $q_{i+2}$},label.side=right}{subpath (length(p2)/6,0) of p2}
  \fmfi{fermion,label={\small $q_{N-1}$},label.side=right}{subpath (4length(p2)/6,3length(p2)/6) of p2}
  \fmfi{fermion,label={\small $p_{N-1}$},tension=2.0}{point 4length(pp2)/6 of pp2 -- point 4length(p2)/6 of p2}
  \fmfi{fermion,label={\small $q_{N}$},label.side=right}{subpath (5length(p2)/6,4length(p2)/6) of p2}
  \fmfiset{p4}{point 5length(pp2)/6 of pp2 -- point 5length(p2)/6 of p2}
  \fmfi{fermion,label={\small $p_{N}$},tension=1.0}{subpath (0,length(p4)) of p4}
  %\fmfi{fermion,label={\small $q_{N+2}$},label.side=right,tension=1.0}{subpath (length(p4)/2,length(p4)) of p4}
  \fmfi{fermion,label={\small $q_{1}$},label.side=left}{subpath (5length(p2)/6,6length(p2)/6) of p2}
  \fmfiset{p5}{point 5length(pp1)/6 of pp1 -- point 5length(p1)/6 of p1}
\fmfi{fermion,label={\small $p_i$},tension=1.0}{subpath (0,length(p5)) of p5}
  %\fmfi{fermion,label={\small $q_{N+3}$},label.side=left,tension=1.0}{subpath (length(p5),length(p5)/2) of p5}
\fmfiset{p3}{point length(p5) of p5 .. a3 .. point length(p4) of p4}
%\fmfi{fermion,rubout,label={\small $q_{N+1}$},label.side=left}{subpath (0,2.6length(p3)/5) of p3}
\fmfi{fermion,rubout,label={\small $q_{N+1}$},label.side=right}{subpath (0,length(p3)) of p3}
  %\fmfi{plain,rubout}{subpath (2.6length(p3)/5,length(p3)) of p3}
  %
\end{fmfgraph*}}
\caption{\small diagram picturing a two-loop $N$-point function with 
a planar topology with four-leg vertices.}
\label{fig8} 
\end{figure}

\noindent
This property of factorisation of $\det(A)$ in the quantities $V$ and $C$ holds for every topology in a three-leg vertex scalar theory. Indeed, it is always possible to choose the internal 4-momenta in such a way that the matrix $\overline{B}$ is homogeneous of degree 1 in the $u_i$'s after the reparametrisation of the Feynman parameters.
Let us conclude this appendix by remarking that this property does not always hold for scalar theory having four-leg vertices. Indeed, there exists some topologies where an internal line connects two four-leg vertices as depicted in fig.~\ref{fig8}. 
\vspace{1cm}
In this case, the $\hr_i$'s are given by:
\begin{align}
\hr_{1} &= \hr_{i} - t_1 &\quad \hr_{i+2} &= \hr_{i+1} -t_{i} \notag \\
\hr_{2} &= \hr_{i} - t_2 &\quad \hr_{i+3} &= \hr_{i+1} -t_{i+1} \notag \\
\hr_{3} &= \hr_{i} - t_3 &\quad \hr_{i+4} &= \hr_{i+1} -t_{i+2} \notag \\
\vdots & & \vdots & \notag \\
\hr_{i-1} &= \hr_{i} - t_i &\quad \hr_{N} &= \hr_{i+1} -t_{N-2} \notag \\
\hr_{N+1} &= \hr_{i} + \hr_{i+1} - t_{N-1} & 
\label{eqfdc2}
\end{align}
with
\begin{align}
  t_{1} &= p_{[1..i-1]} &\quad t_{i} &= p_{i+1} \notag \\
  t_{2} &= p_{[2..i-1]} &\quad t_{i+1} &= p_{[i+1..i+2]} \notag \\
  t_{3} &= p_{[3..i-1]} &\quad t_{i+2} &= p_{[i+1..i+3]} \notag \\
  \vdots & & \vdots & \notag \\
  t_{i-1} &= p_{i-1} &\quad t_{N-2} &= p_{[i+1..N-1]} \notag \\
  t_{N-1} &= -p_i &
  \label{eqfdc21}
\end{align}
The three sets of internal line labels become in this case:
\begin{align}
  S_1 &= \{1, \cdots,i\}, \quad S_2 = \{i+1, \cdots, N\} \quad \text{and} \quad S_3 = \{N+1 \} 
  \label{eqdefsetc}
\end{align}
It is easy to realise that the $\overline{B}$ matrix will not depend on $\tau_i$ and $\tau_{i+1}$ but will depend on $\tau_{N+1}$. For this reason, after the reparametrisation of the Feynman parameters, the $\overline{B}$ matrix will not be homogeneous of degree 1 in the $u_i$'s and thus the factorisation of $\det(A)$ in the $V$ and $C$ terms does not hold.

%In fact, this factorisation property will not hold in the case of a theory with four-leg vertices because there will be Feynman diagram where all the internal legs are connected to vertices also connected to external legs or, in other words, the number of internal lines is equal $N-1$. In this case, it is not possible to have for each set one internal line which are only parametrised by the two $\hr$.

%\input{bibliography}
\bibliographystyle{unsrt}
\bibliography{../biblio,../publi}

 \end{fmffile}
\end{document}